\documentclass[titlepage,12pt]{utarticle}
\usepackage{amssymb,amscd}
\usepackage{floatflt,graphicx} 

\numberwithin{equation}{section}

\newcommand{\ket}[1]{\lvert #1\rangle}
\newcommand{\kket}[1]{\lvert #1\rangle\rangle}

\newcommand{\bl}{\overline{\lambda}}

\newcommand{\sbullet}{\cdot}
\newcommand{\tr}{{\rm tr}}

\begin{document}
\preprint{
  UTTG--01--01\\
  RUNHETC--2000--57\\
  {\tt hep-th/0102018}\\
}
\title{
  Torsion D-Branes\vphantom{y}\\
  in Nongeometrical Phases
}
\author{
  Ilka Brunner
    \thanks{Work supported in part DOE Grant DE-FG02-96ER40959.
    }
    \address{
     Department of Physics and Astronomy\\
     Rutgers University\\
     Piscataway, NJ 08855 USA\\
     {~}\\
     \email{ibrunner@physics.rutgers.edu}
    }
  and Jacques Distler
    \thanks{Work supported in part by NSF Grant PHY0071512
      and the Robert A.~Welch Foundation.
    }
    \address{
      Theory Group, Physics Department\\
      University of Texas at Austin\\
      Austin, TX 78712 USA\\
      {~}\\
      \email{distler@golem.ph.utexas.edu}
    }
}
\date{January 31, 2001}

\Abstract{
In a geometrical background, D-brane charge is classified by topological
K-theory. The corresponding classification of D-brane charge in an arbitrary,
nongeometrical, compactification is still a mystery. We study D-branes on
non-simply-connected Calabi-Yau 3-folds, with particular interest in the
D-branes whose charges are torsion elements of the K-theory. We argue that
we can follow the D-brane charge through the nongeometrical regions of
the K\"ahler moduli space and, as evidence, explicitly construct
torsion D-branes at the Gepner point in some examples. In one of our
examples, the Gepner theory is a nonabelian orbifold of a tensor product
of minimal models, and this somewhat exotic situation seems to be
essential to the physics.
}

\maketitle


\section{Introduction}\label{sec:intro}

The D-branes that were analyzed  thus far are mainly BPS branes
preserving a fraction of the space-time supersymmetry. Starting
with the work of Sen, it has been realized that there are also
stable but non-BPS D-branes. The existence
of (some of) these states has then been interpreted by Witten:
D-brane charges 
are classified by K-theory \cite{Minasian-Moore:K-Theory,WittK} rather than
cohomology. In this context, Sen's non-BPS
states are stable because they carry a conserved discrete charge -- a 
torsion element of the K-theory\footnote{Not all stable, non-BPS 
D-branes are explained this way. For instance, among the D-branes of 
\cite{BerGab,Senrev} are ones which are stable in some
\emph{region} of the moduli 
space, for energetic reason, not because they carry a conserved 
charge.}. There are two possible 
approaches to understand the physics of these stable non-BPS states:
First, one can use a microscopic formulation, where
D-branes are described as boundary conditions in a conformal
field theory. Alternatively, one can use the language
of K-theory and vector bundles. Both of these approaches
have been used to study toroidal orbifolds in the literature.
In this paper, we study non-BPS D-branes in a geometrically
more interesting background -- Calabi-Yau compactifications
whose K-theory has a torsion part. We are particularly
interested in the dependence of the physics of torsion D-branes
on the moduli of the closed string background.

In the large-radius limit, where geometrical reasoning applies, we 
know that D-brane charge is classified by topological K-theory. 
These geometrical concepts are no longer available in the
stringy regime. Here, boundary conformal
field theory provides a powerful tool for the investigation of
stringy D-brane physics. While these methods are in principle
applicable at generic points in moduli space, they are most
useful at rational or ``Gepner'' points, where the theory is
exactly solvable.

Still, while a detailed description of the D-branes may not be
possible at a generic 
point in the moduli space, one might hope to achieve a cruder goal, 
namely the classification of the allowed D-brane charges. That is, we
would like to define a ``quantum K-theory''
(the phrase appeared in \cite{DiaGom}, in analogy with quantum 
cohomology) which would classify the allowed D-brane charges 
everywhere in the moduli space, and which would reduce to topological 
K-theory in the geometrical limit. And, wherever possible, we would 
like to make contact between it and the results of boundary conformal 
field theory.

Turning on a (flat) B-field leads rather naturally to a twisted version 
 of the differential K-theory of 
\cite{Freed:DifferentialCohomology,Freed-Hopkins:RR} (see also \cite{Kapu, 
BouMat} for earlier work). But to define our quantum K-theory, we would 
also need to understand how it deforms when one turns on a finite value for the
K\"ahler modulus. 

While we won't be able to provide a definition of quantum K-theory, 
we will, at least, be able to describe how the quantum K-groups vary 
as we vary the moduli. More precisely, we will be able to determine 
the automorphism of the quantum K-theory that results from traversing 
an incontractible cycle in the moduli space.

For the free part of the K-theory, these automorphisms are calculable 
using Mirror Symmetry. The comparison of BPS D-brane charges
at different points in moduli spaces was studied in
\cite{BDLR, DiaRom, KLLW, Sche, GovJayI, GoJaSaI, DiaDou, Mayr, Toma, GovJayII}.
Our challenge is to extend these results to 
include the torsion in the K-theory. In the simplest class of 
examples (a more precise criterion will be discussed in 
\S\ref{sec:Pairings}), the automorphisms all act trivially on the 
torsion subgroup. So the torsion subgroup of the quantum K-theory is, 
in some sense, ``constant" and independent of the moduli.

We won't address the following much 
harder question. Given a (torsion) class in 
the quantum K-theory at a point in the moduli space: is there a 
stable D-brane in the conformal field theory at that point 
representing the given quantum K-theory class? The analogous question 
for BPS branes is hard enough \cite{DoFiRoI,
DoFiRoII,Douglas:DerivedCategories}.

We \emph{will} see that, in some examples, it is possible to 
construct a stable brane in terms of boundary conformal field theory
at the Gepner point, corresponding to a torsion 
class in the K-theory. This provides some evidence that there is, at 
least, a path between large-radius and the Gepner point along which 
the torsion branes are stable. It is not so clear that the torsion 
branes are stable near the conifold point (see \S\ref{KofX}).

We will present two examples: Both of them are quotients
of the quintic by a freely-acting discrete group. The first is a 
quotient by a freely-acting $\BZ_5$ scaling symmetry. The K\"ahler 
moduli space has three distinguished points: a ``large-radius'' point, 
a Gepner point, and a (mirror of a) conifold point. We construct 
torsion branes at large-radius and as bound states of BPS branes at 
the Gepner point. We find explicitly that the monodromies about these two 
points act trivially on the torsion subgroup of the quantum K-theory. 
Since there are only three boundary points of the K\"ahler moduli 
space, the monodromy about the conifold point must also act trivially 
on the torsion subgroup.  

In the second example, we mod out by an additional orbifold action,
which is a cyclic permutation of the five coordinates. This
example was investigated by Aspinwall and Morrison \cite{AspMor}. 
They found that the tree-level topological string amplitudes (the 
chiral ring) have a $\BZ_5$ symmetry, relating different points in the 
moduli space, which is not a symmetry of the full theory. In 
particular, the 1-loop topological string amplitudes distinguish
these points. We find that the D-brane spectrum is also not invariant 
under this would-be $\BZ_5$ symmetry. Rather, the CFTs at points in 
the moduli space related by this $\BZ_5$ differ by turning on (topologically
trivial) discrete torsion. This, of course, alters the 
spectrum of D-branes. We determine the monodromies (up to a certain ambiguity,
involving the torsion) in the K\"ahler moduli space  and, in particular, we
determine which D-branes become massless at each of the five  points
corresponding to (mirrors of a) ``conifold''-type singularity.

Section \S\ref{sec:methods} is devoted to laying out our guiding
assumptions in determining the monodromies in the quantum K-theory as one
moves about in the moduli space; subsections
\S\ref{sec:charge},\S\ref{sec:PushForward} and \S\ref{sec:AHSS}
contain essential mathematical background to our computations.
Sections \S\ref{sec:BCFTgeneral},\S\ref{sec:Gepner} review the essential
conformal field theoretic background. In \S\ref{sec:GepnerBPS}, we
construct the BPS boundary states on the Gepner orbifolds we are
interested in and in \S\ref{sec:GepnerNonBPS}, we superpose them to
form the desired torsion branes. \S\ref{sec:ExampleX} and \S\ref{sec:W}
are devoted to our examples, and a detailed comparison of the conformal
field theoretic and geometrical results.

\section{K-theory and D-branes}\label{sec:methods}
We begin  with a short review of some relevant facts about K-theory. Much of
this material will be familiar to many readers; we include it here for
completeness.

\subsection{D-brane charge}\label{sec:charge}
In this paper we will be considering Type II string theory on spacetimes of
the form $\BR^4\times X$, where $X$ is a compact Calabi-Yau manifold. We will
be interested in wrapped D-branes which correspond to particles in the
noncompact $\BR^4$. As has become familiar, D-brane charge takes values
in K-theory, but precisely which K-theory is relevant here? For D-branes
which correspond to particles in $\BR^4$, this is
$K^\bullet_{\widehat{cpt}}(\BR^4\times X)$, where by
$K_{\widehat{cpt}}$, we mean  compactly-supported in space, but
not in time. Since the time-direction is contractible, this is
\begin{equation*}
  K^\bullet_{\widehat{cpt}}(\BR^4\times X)=K^\bullet_{cpt}(\BR^3\times X)
\end{equation*}
We can identify $K^\bullet_{cpt}(M)=K^\bullet(M,\partial M)$, the relative
K-theory of $M$ with respect to its boundary.
In our case, $\partial(\BR^3\times X)=S^2\times X$. The relative K-theory
fits into a 6-term exact sequence
\begin{equation}\label{eq:relative}
\begin{CD}
  K^0(M)       @>>> K^0(\partial M) @>>> K^1(M,\partial M)\\ 
  @AAA         &           &    @VVV \\
  K^0(M,\partial M) @<<< K^1(\partial M) @<<< K^1(M)  
\end{CD}
\end{equation}
To evaluate the various terms in this sequence, we use the fact that if
$K^\bullet(X)$ or $K^\bullet(Y)$ is freely-generated, there is a K\"unneth
formula for $K^\bullet(X\times Y)$,
\begin{equation}
\begin{split}
K^0(X\times Y)&=K^0(X)\otimes K^0(Y)\oplus K^1(X)\otimes K^1(Y)\\
K^1(X\times Y)&=K^0(X)\otimes K^1(Y)\oplus K^1(X)\otimes K^0(Y)
\end{split}
\end{equation}
In our case, $K^0(\BR^3)=\BZ$, $K^0(S^2)=\BZ+\BZ$ and
$K^1(\BR^3)=K^1(S^2)=0$.  Plugging this information into \eqref{eq:relative},
we  have
\begin{equation}\label{eq:ourrelative}
\begin{CD}
    K^0(X)       @>i^{*}>> K^{0}(S^{2})\otimes K^0(X) @>>> K^1(\BR^{3}\times 
  X,S^{2}\times X)\\ 
  @AAA         &           &    @VVV \\
  K^0(\BR^{3}\times X,S^{2}\times X) @<<< K^{0}(S^{2})\otimes K^1(X) 
  @<i^{*}<< K^1(X)  
\end{CD}
\end{equation}
The maps $i^{*}$ are injective; they involve taking a class on 
$\BR^{3}$
and restricting it to the 2-sphere at infinity. So $K^1(\BR^{3}\times 
  X,S^{2}\times X)=K^{0}(S^{2})\otimes K^0(X)/i^{*}(K^0(X))=K^0(X)$ and
similarly for $K^0$. So, for Type IIB, our D-brane charge takes values in 
\begin{equation}\label{eq:shift0}
K^0_{\widehat{cpt}}(\BR^4\times X)\simeq K^1(X)
\end{equation}
and for Type IIA, it takes values in
\begin{equation}\label{eq:shift1}
K^1_{\widehat{cpt}}(\BR^4\times X)\simeq K^0(X)
\end{equation}

This shift in degree should be familiar from ordinary E\&M.
There, the charge density, $j\in {\mathrm H}^3_{\widehat{cpt}}(M_3\times \BR)
={\mathrm H}^3_{cpt}(M_3)$ and by a similar exact sequence for relative cohomology,
we find
${\mathrm H}^3_{cpt}(M_3)\simeq \coho{2}{\partial M_3}/i^*(\coho{2}{M_3})$. This isomorphism
is known as \emph{Gauss's Law}; we compute the charge by integrating a 2-form,
$*F$, over the boundary of a region of space. So we should regard
\eqref{eq:shift0},\eqref{eq:shift1} as the K-theoretic analogues of Gauss's Law.

\subsection{The push-forward}\label{sec:PushForward}

We will be particularly interested in the K-theory classes corresponding to
D-branes wrapped around submanifolds of $X$. Indeed, these will form a basis
for the K-theory. The basic construction we will need is the K-theoretic
push-forward. Given a submanifold $i: S\hookrightarrow X$, and a K-theory
class $v$ on $S$, we can use the Thom homomorphism in K-theory to obtain a
K-theory class, $i_!v$, on $X$. This class is characterized by the
Atiyah-Hirzebruch Theorem.

Let $f:Y\to X$ be a continuous map between smooth, compact, connected manifolds.
If $dim(Y)-dim(X)=0 \mod 2$, then for each $a\in K^0(Y)$, there is a
class $f_!a\in K^0(X)$ such that
\begin{equation}\label{eq:AtiyahHirzebruchThm}
ch(f_!a)\hat A(X)= f_* \left(ch(a)e^{{1\over2}d}\hat A(Y)\right)
\end{equation}
where $d\in \coho{2}{Y}$ is a class whose mod-2 reduction is $w_2(Y)-f^*w_2(X)$
and $f_*$ is the push-forward in cohomology. The definition of the
push-forward, $f_!$,
\emph{depends} on a choice\footnote{
  There is an obstruction to defining the push-forward if no such class, $d$,
  exists. The implications of this for D-branes was explored in
  \cite{Freed-Witten}.
} of the class $d$. If $Y,X$ are almost complex manifolds, there is a canonical
choice for $d$: set\footnote{
  This choice of $d$ implements the twisting of
  \cite{Bershadsky-Sadov-Vafa}.
} $d=c_1(Y)-f^*c_1(X)$
and \eqref{eq:AtiyahHirzebruchThm} simplifies to
\begin{equation}\label{eq:AtiyahHirzebruchThmSimple}
ch(f_!a)Td(X)=f_*(ch(a)Td(Y))
\end{equation}
If $dim(Y)-dim(X)=1\mod 2$, then $f_!a\in K^1(X)$. Viewing $K^1(X)$ as a
subgroup of $\tilde K^0(X\times S^1)$, we have the same formula
\eqref{eq:AtiyahHirzebruchThm}, with $X$ replaced by $X\times S^1$ (so
the difference in dimensions is again even).

In the presence of torsion, \eqref{eq:AtiyahHirzebruchThmSimple} (and 
similar formul\ae\ that we use elsewhere) may 
not completely characterize the push-forward. In the dimension we are
working, we can multiply by 6 to clear denominators. If (as will be
true in our examples) there are no elements which are 
2-torsion or 3-torsion, this operation has no kernel. The resulting
equality would, in general, hold as an equality of integral Chern
classes only \emph{modulo torsion}. But, in our application, $Y$ is a
complex submanifold of a Calabi-Yau 3-fold, $X$. The push-forward in
K-theory is supported in a tubular neighbourhood of $Y$ For complex
codimension 2 or 3, the compactly-supported K-theory of the tubular neighbourhood (which is isomorphic to the K-theory of $Y$) is
torsion-free, so the equality actually holds over the integers.
In complex codimension 1 (Y a divisor on $X$), we can use the simpler
result that $i_!\CO_Y=L-\CO$, where $L$ is the line
bundle whose divisor is $Y$.

\subsection{Monodromies}\label{sec:Monodromies}
We are, of course, not just interested in compactification on a particular
Calabi-Yau manifold, $X$, but on a family of such manifolds, parametrized by
some moduli space, $\CM$. Over each point in $\CM$, the allowed D-brane charges
form a discrete abelian group, $\Gamma$.  Locally, there is not much structure;
since the charges are discrete, not much can happen to them. But if we circle
a singular locus (the singular loci are complex-codimension one in $\CM$), the
group $\Gamma$ will come back to itself only up to an automorphism. So we
have a $\Gamma$-bundle over the moduli space, and this bundle is characterized
by a set of automorphisms $\in Aut(\Gamma)$, one for each of the for each of
the homotopically-nontrivial paths in $\CM$.

If we mod out by the torsion in the K-theory, the quotient $K(X)/K(X)_{tor}$
is a lattice and we have more structure. If we map $K(X)\to K(X,\BC)$,
then the latter forms a complex (in fact, holomorphic) vector bundle over
$\CM$, with a flat connection (whose flat sections may be taken to be
$K(X)/K(X)_{tor}$). The automorphisms of the lattice are just given by the
holonomies of this connection and can be written as monodromy matrices with
respect to a basis of flat sections. These are essential ingredients in defining
the Special Geometry of $\CM$.

On the level of $K(X)/K(X)_{tor}$, the monodromies may be determined, rather
explicitly, using Mirror Symmetry. Our task, however, will be to
extend them to act on all of $K(X)$, including the torsion. For this purpose, it will
prove more insightful to write topological (K-theoretic) formul\ae.
For this, we follow \cite{Kont,Horj}.

We will mostly study the monodromies in the K\"ahler moduli space, and at
that, we will mostly specialize to the case of one-dimensional K\"ahler
moduli spaces. The monodromy around the large-radius limit is given shifting
the $B$-field by the generator, $\xi$, of $\coho{2}{X}$. Since the B-field enters
into the Chern character by
\begin{equation}
 Ch(\CF)=Tr\left(e^{F+B}\right)
\end{equation}
we see that shifting $B\to B+\xi$ corresponds to tensoring with a line
bundle, $L$ such that $c_1(L)=\xi$.

Another singular locus  of the moduli space is the (mirror of the) conifold,
at which the volume of $X$ shrinks to zero size (while the volumes of
2-cycles and 4-cycles stay finite) \cite{Greene-Kanter}. In this case, the
monodromy is \cite{Kont,Horj}
\begin{equation}\label{eq:ConifoldMonodromy}
v\mapsto v- k Ind(\overline{\partial}_v)\CO
\end{equation}
Here $v$ is an element of $K^0(X)$, and $\CO$ is the trivial line bundle.
In the case of hypersurfaces in toric varieties (as considered by \cite{Horj}),
the coefficient $k$ is 1. For the orbifolds we will consider later, we will
find $k>1$.

\subsection{Pairings}\label{sec:Pairings}
If all we had was the structure of an abelian discrete group (modulo
torsion, a lattice), then the automorphisms could be rather more general.
However, there are bilinear pairings on the K-theory, and we must demand
that the automorphisms be symmetries of these pairings.

First, there is the usual intersection pairing,
\begin{equation}
(.,.): K^0(X)\times K^0(X) \to \BZ
\end{equation}
This is a skew-symmetric bilinear pairing given by
\begin{equation}
    (v,w)=Ind(\overline{\partial}_{v\otimes\overline{w}})
      =\int_X ch(v\otimes\overline{w})Td(X)
\end{equation}
Normally, this is written in terms of the Dirac Index, rather than
the Dolbeault Index. They agree on a Calabi-Yau manifold. 
It is skew-symmetric because
$Ind(\overline{\partial}_{\overline{v}})=-Ind(\overline{\partial}_{v})$. It
clearly annihilates $K^0_{tor}(X)$ and it is nondegenerate on
$K^0(X)/K^0_{tor}(X)$.

There is a similar pairing on $K^1(X)$, given by the natural skew-symmetric map
\begin{equation*}
K^1(X)\times K^1(X) \to K^0(X)
\end{equation*}
(where we have used Bott periodicity), followed by taking the index. Again,
this is nondegenerate on $K^1(X)/K^1_{tor}(X)$.

These intersection pairings annihilate the torsion in K-theory. To capture
information about the torsion, we can study the torsion-pairing
\cite{AtPaSiI,AtPaSiII,AtPaSiIII}. This is a little more subtle to define, and we
will review its definition and the relation to conformal field theory 
elsewhere \cite{Brunner-Distler:WIP}. For present purposes, it suffices
 that it is a nondegenerate pairing
\begin{equation}
\langle.,.\rangle: K^0(X)_{tor}\times K^1(X)_{tor} \to \BR/\BZ
\end{equation}
In our examples (this is \emph{not} the general situation!), 
$K^{1}(X)_{tor}$ is generated by D1-branes wrapped around torsion 
1-cycles on $X$\footnote{In general, these generate only a subgroup of
$K^1(X)_{tor}$, and similarly for $K^0(X)_{tor}$.}, and
$K^{0}(X)_{tor}$ is generated by elements of the form $L-\CO$, for $L$ a flat
line  bundle on $X$. In this case, the torsion pairing is just the holonomy 
of the flat line bundle (corresponding to an element of
$K^{0}(X)_{tor}$) around the torsion 1-cycle (corresponding to an
element of $K^{1}(X)_{tor}$).

Now, it is a fundamental feature of special geometry that the 
monodromies on the K\"ahler moduli space leave $K^{1}(X)/K^{1}(X)_{tor}$ 
invariant, and similarly, the monodromies on the complex-structure 
moduli space leave $K^{0}(X)/K^{0}(X)_{tor}$ invariant. So, if we 
ignore the torsion, we need only check the invariance of one or the 
other of the intersection pairings.

With the torsion, things are more subtle. If a monodromy on the K\"ahler 
moduli space acts as a nontrivial automorphism on $K^{0}(X)_{tor}$, 
then, for the torsion pairing to be invariant, it must \emph{also} act 
nontrivially on $K^{1}(X)_{tor}$ while leaving $K^{1}(X)/K^{1}(X)_{tor}$ 
invariant. 

In our examples, we will, in fact, find that all of $K^{1}(X)$, including the
torsion, is actually 
invariant under the monodromies of the K\"ahler moduli space, or 
equivalently that $K^{0}(X)_{tor}$ is invariant. At least for the example of
\S\ref{sec:ExampleX}, that  will be justified \emph{a posteriori} by our
construction of the torsion  D-branes at the Gepner point. For the example of
\S\ref{sec:W}, we were unable to \emph{prove} that the action on the torsion
subgroup was trivial, but we were unable to find a nontrivial action consistent
with all of our other requirements.

As we said, the monodromies must leave the intersection pairing
invariant. That is,
\begin{equation}
    (Mv,Mw)=(v,w)
\end{equation}
This is obviously true for the large radius monodromy, since
\begin{equation}
    (v\otimes L,w\otimes L)=(v,w)
\end{equation}
for any line bundle $L$.

We also need to consider the effect on the torsion pairing.  In the 
examples we consider, the ring structure of the K-theory will be such 
that, if $\alpha\in K^{0}(X)_{tor}$, then $\alpha\otimes L=\alpha$ for 
any line bundle $L$, so that the monodromy acts trivially on $K^{0}(X)_{tor}$.
If we take the monodromy to act trivially on $K^{1}(X)$, we then 
have that both torsion subgroups are fixed by the monodromy, and hence 
(trivially) so is the torsion pairing. More interesting is  a case 
in which $\alpha\otimes H=\alpha'\neq\alpha$. In that case, where the 
monodromy acts as a nontrivial automorphism of $K^{0}(X)_{tor}$.
In order for the torsion pairing to be invariant, the monodromy would 
also have to act nontrivially on $K^{1}(X)_{tor}$ (while still acting 
trivially on $K^{1}(X)/K^{1}(X)_{tor}$).

Similarly the monodromy about the conifold
\eqref{eq:ConifoldMonodromy} leaves the intersection pairing invariant. 
One can see this by writing the monodromy as 
\begin{equation}\label{eq:newConifoldMonodromy}
v\mapsto v- k (v,\CO)\CO
\end{equation}
and using the skew-symmetry of the pairing.

This monodromy has a very simple interpretation. Let's say that at
some singularity, a brane (a charged particle in the 4D effective theory)
corresponding to K-theory class $w$ becomes massless. Circling the singularity
shifts the $\theta$-angle
-- of the $U(1)$ for which this particle is electrically-charged -- by $2\pi$.
By the Witten effect, the charge of a particle in K-theory class $v$ gets
shifted by\footnote{Strictly speaking, this argument involving charges 
only fixes the action on $K^{0}(X)/K^{0}(X)_{tor}$. However, the
monodromy must also commute with the action of the quantum symmetry,
to be discussed in \S\ref{sec:QuantSym}. This, together with the
action on the quotient, will frequently fix the actionon all 
of $K^{0}(X)$. Since this monodromy acts trivially on $K^0(X)_{tor}$, the
torsion pairing is preserved if it also acts trivially on $K^{1}(X)$.} 
\begin{equation*}
 v\mapsto v-  (v,w)w
\end{equation*}
when you shift $\theta\to\theta+2\pi$. If $k$ such particles become massless,
then the shift is \eqref{eq:newConifoldMonodromy}. More generally, if several,
mutually-local, particles become massless, the shift is
\begin{equation}\label{eq:GeneralConifoldMonodromy}
  v\mapsto v-  \sum_i(v,w_i)w_i
\end{equation}
where mutual locality means $(w_i,w_j)=0$. As we mentioned in
\S\ref{sec:Monodromies}, at the conifold, it is the D6-brane, corresponding
to the K-theory class $\CO$ which becomes massless.

\subsection{The Atiyah-Hirzebruch and Cartan-Leray spectral sequences}\label{sec:AHSS}

In order to carry through our computations, we will need to be able to
compute the cohomology of a Calabi-Yau manifold $X=Y/G$, given the knowledge
of the cohomology of the covering space $Y$. And, given the cohomology of $X$,
we will need to be able to compute its K-theory.

Two spectral sequences come to our aid here. The Cartan-Leray spectral
sequence allows one to compute the \emph{homology} of $X$. The Atiyah-Hirzebruch
spectral sequence allows one to compute its K-theory.
Both of these are spectral sequences of a double-complex \cite{Bott-Tu}.

For a \emph{cohomology} spectral sequence (like AHSS), the $r^{th}$
approximation to the cohomology we are after is a bigraded complex,
$E^{p,q}_r$, together with a differential
\begin{equation*}
   d_r: E^{p,q}_r\to E^{p+r,q-r+1}_r
\end{equation*}
which increases the total degree, $n=p+q$, by one. The $(r+1)^{st}$
approximation, $E_{r+1}$ is the cohomology of $d_r$ on $E_r$. Raoul
Bott has likened spectral sequences to perturbation theory, each successive
approximation getting closer and closer to the desired answer. In most
applications, the $E^{p,q}_r$ vanish outside of some finite range in
$p$ (or outside of some finite range in $q$). Consequently, unlike
perturbation theory, the spectral sequence is guaranteed to converge after a
finite number of steps to something we can call
$E^{p,q}_\infty$. What the spectral sequence converges to, however, is not
quite the cohomology we are interested in. Rather, it converges to the
\emph{associated graded},
$Gr({\mathrm H}^{\bullet})$. That is, the cohomology group ${\mathrm H}^n$ has a filtration
(sequence of subgroups, $F^{(n)}_i$)
\begin{equation}
   {\mathrm H}^n=F^{(n)}_0\supset F^{(n)}_1\supset
     F^{(n)}_2\supset F^{(n)}_3\supset\dots
\end{equation}
such that $F^{(n)}_p/F^{(n)}_{p+1}= E^{p,n-p}_\infty$. So, in general,
even after we have computed $E_\infty$, we still have an extension problem
to solve, in order to recover the cohomology groups ${\mathrm H}^n$ themselves.

A \emph{homology} spectral sequence (like CLSS) is very similar. The $r^{th}$
approximation is a bigraded complex, $E^r_{p,q}$, but this time the
differential \emph{lowers} the total degree by one
\begin{equation*}
   d_r: E_{p,q}^r\to E_{p-r,q+r-1}^r
\end{equation*}
Again, the spectral sequence converges to the associated graded
$Gr({\mathrm H}_{\bullet})$, where the homology group ${\mathrm H}_n$ has a
descending filtration
\begin{equation}
   {\mathrm H}_n=F_{(n)}^n\supset F_{(n)}^{n-1}\supset
     F_{(n)}^{n-2}\supset F_{(n)}^{n-3}\supset\dots
\end{equation}
with $F_{(n)}^p/F_{(n)}^{p-1}= E^{p,n-p}_\infty$.

It is conventional to denote the complexes $E_r$ (or $E^r$) by rectangular
arrays, with the index $p$ running in the horizontal direction and the
index $q$ running in the vertical direction. So the differential acts down
and to the right for a cohomology spectral sequence. It acts up and to the
left for a homology spectral sequence.

\paragraph{The Cartan-Leray spectral sequence:}
For a discrete group, $G$, the Eilenberg-MacLane space, $K(G,n)$ has
\begin{equation}
   \pi_p(K(G,n))=
   \begin{cases}
      \BZ&p=0\\
        G&p=n\\
        0&\text{otherwise}
   \end{cases}
\end{equation}
(For $n>1$, $G$ must be abelian.) The group homology of $G$ may be defined
as the ordinary homology of the corresponding Eilenberg-MacLane space
\begin{equation*}
\homo{n}{G}\equiv \homo{n}{K(G,1)}
\end{equation*}
So, we have, immediately that
\begin{equation*}
\homo{0}{G}=\BZ,\qquad \homo{1}{G}=G/[G,G]
\end{equation*}

For any abelian discrete group, $G$, $K(G,1)$ can be determined from
\begin{equation*}
    \begin{split}
        K(\BZ,1)&=S^{1}  \\
        K(\BZ_{n})&=L(\infty,n)=S^{\infty}/\BZ_{n}  \\
        K(G\times G',1)&=K(G,1)\times K(G',1)
    \end{split}
\end{equation*}
A relatively short computation (see, {\it e.g.}~\cite{Brown-GroupCohomology})
yields
\begin{equation}
\homo{2}{\BZ_m\times \BZ_n}=\BZ_{(m,n)}
\end{equation}

If $G$ acts freely on $Y$, the Cartan-Leray spectral sequence allows one to
compute the homology of $X=Y/G$. The $E^2$ term is 
\begin{equation}
E^2_{p,q}= \homo{p}{G,\CH_q(Y)}
\end{equation}
the homology with \emph{twisted coefficients} (see
  \cite{Brown-GroupCohomology} for precise definitions).
If $G$ acts trivially on $\homo{q}{Y}$, the coefficient group is constant:
$\CH_q(Y)=\homo{q}{Y}$, and the homology groups $\homo{p}{G,\homo{q}{Y}}$ are
determined from $\homo{\bullet}{G}$ and the Universal Coefficients Theorem.
If $G$ does not act trivially on $\homo{q}{Y}$, then these homology groups with
twisted coefficients are a bit ugly to compute. The one easy case is
$\homo{0}{G,\CH_q(Y)}$.
Let $\homo{q}{Y}_G$ be the \emph{coinvariant
quotient}, $\homo{q}{Y}_G= \homo{q}{Y}/\CA$, where $\CA$ is the subgroup of
$\homo{q}{Y}$ generated by elements of the form $x-g\cdot x$, for
$x\in \homo{q}{Y}$ and $g\in G$.  One finds
\begin{equation}
   \homo{0}{G,\CH_q(Y)}=\homo{q}{Y}_G
\end{equation}
If $G$ acts trivially on the homology, then
$\homo{q}{Y}_G=\homo{q}{Y}$.

The Cartan-Leray spectral sequence converges to $Gr(\homo{\bullet}{Y/G})$.
Our basic application will be the following. Assume
\begin{itemize}
\item $Y$ is simply connected.
\item $G$ acts freely on $Y$.
\item $\homo{2}{Y}_G$ is torsion-free.
\end{itemize}
The $E^2$ term of the CLSS looks like.
\begin{subequations}\label{eq:CLSS}
\begin{center}\qquad\qquad $E^{2}_{p,q}$
    \begin{tabular}{c|ccccc}
          & $\vdots$ &        &        &        & \\
         2&$\homo{2}{Y}_G$&&&      & \\
         1&    $0$   &  $0$   &  $0$     &        & \\
         0&   $\BZ$  &$\homo{1}{G}$&$\homo{2}{G}$&$\homo{3}{G}$&$\dots$\\
        \hline
          & 0 & 1 & 2 & 3 & 
    \end{tabular}
\hfill\eqref{eq:CLSS}
\end{center}
\end{subequations}
At least for this portion of the diagram, $d_2$ vanishes identically,
so $E^3=E^2$. The differential
$d_3:\homo{3}{G}\to \homo{2}{Y}_G$ could, in principle,
be nontrivial, but vanishes in this case because $\homo{2}{Y}_G$ is
torsion-free\footnote{If $\homo{2}{Y}_G$ has 
torsion, then $d_{3}:\homo{3}{G}\to\homo{2}{Y}_G$ kills part of the 
torsion in $\homo{2}{Y}_G$, and the spectral sequence converges at the 
$E^{4}$ term.}. The spectral sequence converges (since the higher differentials
would land outside of this rectangle). So
\begin{equation}\label{eq:H1}
\homo{1}{X}=\homo{1}{G}=G/[G,G]
\end{equation}
and $\homo{2}{X}$ has the filtration
\begin{equation*}
\homo{2}{X}= F_2\supset F_1\supset F_0
\end{equation*}
where
\begin{equation}\label{H2filt}
   F_0=F_1=\homo{2}{Y}_G,\qquad F_2/F_1=\homo{2}{G}
\end{equation}
This can be rewritten more succinctly as the short exact sequence
\begin{equation}\label{eq:H2extensionnew}
0\to \homo{2}{Y}_G\xrightarrow{\pi_{*}} \homo{2}{X}\to \homo{2}{G}\to 0
\end{equation}
where $\pi_*$ is the push-forward by the projection $\pi:Y\to X$
and makes sense on the quotient, $\homo{2}{Y}_{G}$, 
because elements of $\CA$ push-forward to zero.

If $\homo{2}{G}=0$, then we find that $\homo{2}{X}$ is just $\homo{2}{Y}_G$ and
is torsion-free. Otherwise, we still have a nontrivial extension problem,
\eqref{eq:H2extensionnew}, 
to solve and $\homo{2}{X}$ may or may
not have torsion.

As a simple application, consider the Tian-Yau manifold, which was the second
example in \cite{Brau}. The K-theory of that manifold has a nontrivial
torsion part, but the author of that paper did not compute it.
The missing ingredient was a computation of the torsion in 
$H_{2}(X)$. With our methods, this proves to be straightforward.
$\homo{2}{Y}$ is 14-dimensional.
Under the action of $G=\BZ_3$, it decomposes as 4 regular representations
and two trivial representations \cite{Distler:27bar-cubed}. For each of
the regular representations (with basis $x_1,x_2,x_3$, which are
cyclically-permuted by $G$), we mod out by the subspace spanned by
$\{x_1-x_2,x_2-x_3\}$.The quotient is torsion-free, and is generated  by
$x_1$ (modulo $\CA$). Since $\homo{2}{G}=0$, we conclude that
$\homo{2}{X}=\homo{2}{Y}_G=\BZ^6$. The considerations below then yield 
$K^{0}(X)_{{tor}}=K^{1}(X)_{tor}=\BZ_{3}$.

\paragraph{The Atiyah-Hirzebruch spectral sequence:}
The Atiyah-Hirzebruch spectral sequence allows one to compute the
K-theory of $X$, given a knowledge of its cohomology. The $E_2$ term is
\begin{equation}
E_2^{p,q}=\coho{p}{X,\pi_q(BU)}
\end{equation}
where $\pi_{2n}(BU)=\BZ$, $\pi_{2n+1}(BU)=0$. The spectral sequence converges to
\begin{equation}
E_\infty^{p,q}=Gr(K^{p+q}(X))
\end{equation}
The key feature of the differentials $d_r$ in the AHSS is their images
are always torsion.

We will be interested in the K-theory of Calabi-Yau manifolds, but the
computation of the AHSS works exactly the same for any compact, connected
6-manifold, $X$, with finite fundamental group (so that $\coho{1}{X}=0$). The
$E_{2}$ term 
looks like
\begin{center} $E_{3}^{p,q}=E_{2}^{p,q}$
    \begin{tabular}{c|cccccccc}
	6&$\coho{0}{X}$ &0& $\coho{2}{X}$ & $\coho{3}{X}$ & $\coho{4}{X}$ & $\coho{5}{X}$ &
$\coho{6}{X}$  \\
        5 &  \multicolumn{6}{l}{
\begin{picture}(50,10)
    \put(30,12){\vector(4,-1){80} 
               \makebox(-100,-8){$\scriptstyle d_3$}
	       }
\multiput(100,16)(45,0){2}{\vector(4,-1){100} 
\makebox(-100,-15){$\scriptstyle d_3$}}\end{picture}}
  &\\
        4&$\coho{0}{X}$ &0& $\coho{2}{X}$ & $\coho{3}{X}$ & $\coho{4}{X}$ & $\coho{5}{X}$ &
$\coho{6}{X}$  \\
        3 &0&0&0&0&0&0&0\\
        2&$\coho{0}{X}$ &0 & $\coho{2}{X}$ & $\coho{3}{X}$ & $\coho{4}{X}$ & $\coho{5}{X}$ &
$\coho{6}{X}$  \\
        1 &0&0&0&0&0&0&0\\
        0&$\coho{0}{X}$ &0& $\coho{2}{X}$ & $\coho{3}{X}$ & $\coho{4}{X}$ & $\coho{5}{X}$ &
$\coho{6}{X}$  \\
        \hline
        & 0 & 1 & 2 & 3 & 4 & 5 & 6 \\
    \end{tabular}
\end{center}

The first potentially nonvanishing differential is $d_3$, since all of the
odd rows of the complex vanish. $d_3$ annihilates $\coho{0}{X}$, since the ``lift"
of $\coho{0}{X}$ to K-theory is just the rank, and there is always a trivial
vector bundle of rank $n$, for any $n$. So the generator of $\coho{0}{X}$ must
survive in the cohomology of $d_3$. Similarly, $d_3$ annihilates $\coho{2}{X}$, since
its ``lift" to K-theory is the first Chern class, and for any $\xi\in \coho{2}{X}$,
we can always find a line bundle $L$ with $c_1(L)=\xi$, so that $L-\CO$ is the
lift to K-theory of $\xi$. Finally, $d_3: \coho{3}{X}\to \coho{6}{X}$ also vanishes
because $\coho{6}{X}$ is torsion-free.

The only possible higher differential, $d_5: \coho{0}{X}\to \coho{5}{X}$ again
vanishes for the same reasons as in the previous paragraph, and so the
spectral sequence converges at the $E_2$ term. $K^0(X)$ has a filtration by
$\coho{ev}{X}$ and $K^1(X)$ has a filtration by $\coho{odd}{X}$.

The filtration on $K^1(X)$ leads to the short exact sequence
\begin{equation*}
    0\to \coho{5}{X}\to K^1(X)\to \coho{3}{X}\to 0
\end{equation*}
Since $\coho{5}{X}$ is torsion, this tells us that the torsion subgroup (which
is what we are really interested in computing) is also given by
\begin{equation}\label{eq:K1ext}
    0\to \coho{5}{X}\to K^1(X)_{tor}\to \coho{3}{X}_{tor}\to 0
\end{equation}
The filtration on $K^0(X)$ is a little longer. Recasting it as a set of
short exact sequences, we have
\begin{subequations}\label{eq:K0filt}
  \begin{gather}
    0\to \tilde K(X)\to K^0(X)\to \coho{0}{X}\to 0\\ 
    0\to \tilde K_{(2)}(X)\to \tilde K(X)\to \coho{2}{X}\to 0\\
    0\to \coho{6}{X}\to\tilde K_{(2)}(X)\to \coho{4}{X}\to 0
  \end{gather}
\end{subequations}
The sequences (\ref{eq:K0filt}a,b) are universal. They single out the reduced K-theory,
$\tilde K(X)$ as the subgroup of $K^0(X)$ with vanishing rank and
$\tilde K_{(2)}(X)$ as the subgroup of the reduced K-theory with vanishing
first Chern class. The sequence (\ref{eq:K0filt}c) is special to our low-dimensional
situation where the AHSS converges at the $E_2$ term. 

The image of $\coho{6}{X}$ in $K^0(X)$ is $i_!\CO_p$, the push-forward of
the trivial line bundle over a point in $X$. This cannot be written as a
multiple of some other K-theory class, so
$\tilde K_{(2)}(X)_{tor}=\coho{4}{X}_{tor}$. Putting that together with
(\ref{eq:K0filt}b), we find that the torsion subgroup is given by the extension
\begin{equation}\label{eq:K0ext}
    0\to \coho{4}{X}_{tor}\to K^0(X)_{tor}\to \coho{2}{X}_{tor}\to 0
\end{equation}

Now, let us put the output of these two spectral sequences together.
Let $Y$ be a Calabi-Yau hypersurface (or complete intersection) in a toric
variety. By the Lefschetz theorem, $\homo{1}{Y}=0$ and $\homo{2}{Y}$ is
torsion-free.
By Poincar\'e duality, this is enough to show that all of $\homo{\bullet}{Y}$ is
torsion-free, and hence so is $\coho{\bullet}{Y}$. Therefore 
$K^{\bullet}(Y)$
is torsion-free.

To obtain something interesting, we mod out by a freely acting finite group to
form $X=Y/G$. For our applications, we will assume that $G$ acts
holomorphically, and preserves the holomorphic 3-form so that $X$ is again
Calabi-Yau, but for these topological considerations, that doesn't matter.
{}From \eqref{eq:H1}, we learn that $\coho{2}{X}_{tor}$ and $\coho{5}{X}$ are
$G/[G,G]$. $\coho{3}{X}_{tor}$ and $\coho{4}{X}_{tor}=\homo{2}{X}_{tor}$ are
determined once we've solved the extension problem \eqref{eq:H2extensionnew} to
find the torsion in $\homo{2}{X}$. Finally, we use
\eqref{eq:K0ext},\eqref{eq:K1ext} to compute the torsion in the K-theory.

\subsection{The quantum symmetry and discrete torsion}\label{sec:QuantSym}
In the \S\ref{sec:AHSS}, we considered Calabi-Yau manifolds of the form $X=Y/G$.
String theory propagating on $Y/G$ is governed by an orbifold conformal 
field theory. Since $G$ acts freely on $Y$, there are no massless states in 
the twisted sectors, just massive string states. Still, the full 
conformal field theory (like any orbifold theory) has a quantum 
symmetry group isomorphic to $G/[G,G]$. How this works in the closed 
string sector is familiar. A little less familiar is how the quantum 
symmetry group acts on the D-branes of the theory.

Under tensor products, the flat line bundles on $X$ form an abelian group
which is isomorphic to $\coho{2}{X}_{tor}=G/[G,G]$. Identifying this with the
quantum symmetry group, it acts on $K^{\bullet}(X)$ by
\begin{equation}\label{eq:QuantumSymmetry}
   v\mapsto v\otimes L
\end{equation}
for $L$ a flat line bundle. Since the quantum symmetry is a symmetry
of this whole \emph{family} of conformal field theories, we obtain
a further constraint on the monodromies discussed in \S\ref{sec:Monodromies}:
they must commute with the action of \eqref{eq:QuantumSymmetry}.

We saw that $\homo{2}{X}_{tor}\neq0$, was the condition for having  a
potentially nontrivial extension problem \eqref{eq:K0ext},\eqref{eq:K1ext}
relating the K-theory to the cohomology. It is also the condition under which
one can turn on a topologically-nontrivial flat B-field, {\it i.e.}~one with
$H\in\coho{3}{X}_{tor}$. Turning on this \emph{discrete torsion} modifies the
spectrum of the closed string theory and the D-brane charges take values in
twisted K-theory, $K^\bullet_H(X)$, where $H$ is the
class in $\coho{3}{X}_{tor}$ \cite{Kapu,BouMat}.
The moduli space of the compactification, in this case, consists of disconnected
components, labeled by the discrete torsion. 

Viewed as an orbifold of the conformal field theory on $Y$, however, the 
theory on $X=Y/G$ admits discrete torsion whenever $\homo{2}{G}$
is nonzero, \emph{even if} $\homo{2}{X}_{tor}$ (and hence 
$\coho{3}{X}_{tor}$) vanishes (see \eqref{eq:H2extensionnew}). In that 
situation, we have a flat B-field, but there is no topological 
obstruction to continuously turning it off. Hence the theory ``with 
discrete torsion'' is continuously connected to the theory ``without'' 
discrete torsion -- they lie in the same connected component of the 
moduli space. We will encounter an example of this in \S\ref{sec:W}.

\subsection{Our examples}
Perhaps the best-known Calabi-Yau Manifold is the 
quintic hypersurface, $Y$, in $\BC P^{4}$. The Fermat quintic,
\begin{equation*}
   z_1^{5}+z_2^{5}+z_3^{5}+z_4^{5}+z_5^{5}=0
\end{equation*}
where the  $z_{i}$ are the homogeneous coordinates on 
$\BP^{4}$, is invariant under a freely-acting $\BZ_{5}\times\BZ_{5}$
symmetry generated by
\begin{equation}\label{eq:FirstZ5}
    (z_{1},z_{2},z_{3},z_{4},z_{5})\to (z_{1},\omega 
    z_{2},\omega^{2}z_{3},\omega^{3}z_{4},\omega^{4}z_{5})
\end{equation}
where $\omega^{5}=1$ and
\begin{equation}\label{eq:SecondZ5}
    (z_{1},z_{2},z_{3},z_{4},z_{5})\to (z_{2},
    z_{3},z_{4},z_{5},z_{1})
\end{equation}
The quintic, of course, is simply connected and its K-theory is torsion-free.
To form a non-simply 
connected Calabi-Yau, we mod out by \eqref{eq:FirstZ5} to form the 
manifold $X=Y/\BZ_{5}$. $X$ will be our first example. It is nice, in 
that we can construct the corresponding orbifolded Gepner model
quite explicitly, and study the boundary states there.

We can go further and mod out by \eqref{eq:SecondZ5} to 
form $W=X/\BZ_{5}$. This will be our second example. Several new 
features will arise.

\section{BCFT Generalities}\label{sec:BCFTgeneral}
Once we move away from the large-radius limit, we have to use the
language of conformal field theory to describe the compactification.
A framework for studying D-branes in such compactifications
is provided by boundary conformal field theory. A conformal
field theory on a Riemann surface with a boundary requires
specifying boundary conditions. For $\sigma$-models these
conditions are given by Dirichlet or Neumann boundary conditions
on the $\sigma$-model fields.
Consider now a general conformal field theory
with a chiral symmetry algebra ${\cal A}$. In this case
a class of boundary conditions is provided by the automorphisms $\Omega$
of the chiral algebra. These can be used as gluing conditions
for the symmetry generators $W$ along the boundary taken to be the real line:
$W(z)=(\Omega \bar W)(\bar z)$, for $z=\bar z$.
A generalization of Cardy's formalism
provides,  in this case,  a set of boundary states, 
see \cite{RecSch,Cardy} for more details.

In the context of Calabi-Yau compactification, the chiral
algebra which has to be preserved along the boundary, is
the $\CN=2$ world-sheet supersymmetry algebra. This algebra has
a well-known automorphism, the mirror-automorphism. Accordingly,
there are two types of boundary conditions \cite{OoOzYi}:
\begin{equation}
     \begin{split}
        T&=\bar T, \quad J=-\bar J, \quad G^{\pm} = \bar G^{\mp}
               \quad \textrm{``A-type''}\\
        T&=\bar T, \quad J=\bar J, \quad G^{\pm} = \bar G^{\pm}
               \quad \textrm{``B-type''}
   \end{split}
\end{equation}
These boundary conditions guarantee that the boundary
theory preserves one copy of an $\CN=2$ algebra.
For the construction of BPS branes in CFT it is required
also that $\CN=1$ spacetime supersymmetry is preserved.
The spacetime supersymmetry generator is obtained from the
spectral flow operator of the worldsheet theory.
Bosonizing the $U(1)$ current $J =i\sqrt{c/3} \  \partial X$,
the spectral flow operator is given by $\exp{(i\sqrt{3/c}\,\eta X)}$,
where $\eta$ specifies the number of units of spectral flow.
To construct a space-time SUSY generator, one half unit
of spectral flow is needed. The obvious boundary conditions
on the spectral flow operator compatible with A-type
or B-type boundary conditions are:
\begin{equation}\label{eq:bozo}
     \begin{split}
        e^{i\sqrt{3/c} \, \eta X_L}&=e^{2\pi i\varphi} 
                \ e^{-i\sqrt{3/c} \, \eta X_R}
              \quad \textrm{``A-type''}\\
        e^{i\sqrt{3/c} \, \eta X_L}&=e^{2 \pi i\vartheta} \ 
                e^{i\sqrt{3/c} \,\eta X_R}
               \quad \textrm{``B-type''}
   \end{split}
\end{equation}
We see that the A-type or B-type  conditions are
Dirichlet or Neumann boundary condition on the boson $X$,
describing the bosonized $U(1)$. Accordingly,  
a ``position'' can be associated with  A-type states and a ``Wilson line''
with  B-type states.
The phase appearing in the boundary conditions
determines which space-time $\CN=1$ algebra is preserved.
For two or more D-branes, the difference in the phase determines
if there is a common preserved supersymmetry \cite{GutSat,Doug}. $\vartheta$ can
be interpreted as a ``Wilson line'' along the
bosonized $U(1)$ for the B-type branes. We will see these
Wilson lines explicitly in the case of the Gepner model BPS
branes. Similarly, the A-type states have positions along
the spectral flow direction.

Far out at large volume, A-type BPS boundary states correspond to branes
wrapping special Lagrangian submanifolds of the Calabi-Yau,
and B-type BPS branes correspond to vector bundles on
holomorphic cycles \cite{OoOzYi}. In K-theory language, A-type branes
are classified by $K^1$, whereas B-type branes are classified
by $K^0$.

For non-BPS states, 
the spectral flow symmetry can be broken by the boundary,
since the brane breaks spacetime
supersymmetry. Consistency does however require that the
worldsheet supersymmetry  is preserved, in other
words, there will be A-type and B-type non-BPS states.
In the examples introduced in the previous section, we expect
to find two types of stable but non-BPS branes, one in $K^0$,
and one in $K^1$. In this paper, we will give a physical 
description of these branes as bound states of BPS D-branes,
similar to the approach taken in \cite{BerGab,GabSte,Senrev, Gabrev}
for orbifolds of tori and flat space.

The aim in the following sections is to identify a stack of BPS D-branes,
described as boundary states in Gepner models,
whose stable ground state is the torsion brane.

\section{Gepner Models and Orbifolds Thereof}\label{sec:Gepner}
The stringy regime of Calabi-Yau orbifolds is described
by orbifolds of Gepner models, whose closed string
sector  is reviewed in this
section. For this, we first recall the Gepner model
and its symmetries and then construct the closed
string partition function of the orbifold model.

\subsection{Building blocks of Gepner models}
The basic building blocks of the Gepner model are
$\CN = 2$ minimal models. The minimal model $MM_k$
has the following coset representation:
\begin{equation}
MM_k = \frac{SU(2)_{k}\times U(1)_4}{U(1)_{2k+4}}
\end{equation}
Accordingly, the central charge is $c_k = 3k/(k+2)$.
The irreducible representations of the theory are labeled by $(l,m,s)$,
where $l$ refers to the $SU(2)$, $m$ to the $U(1)_{2k+4}$
in the numerator and $s$ to the $U(1)_4$ in the nominator.
These three integers are subject to an additional constraint:
\begin{equation*}
     l+ m+ s \quad even.
\end{equation*}
The symmetry group of a minimal model is
$\BZ_{2k+4} \times \BZ_2$ for $k$ even and $\BZ_{4k+8}$
for $k$ odd. It is generated by operators with the
following action on primary fields:
\begin{equation}\label{sym}
     \begin{split}
       g \ \Phi_{(l,m,s)} &= e^{2\pi i \frac{q}{k+2}} \ \Phi_{(l,m,s)} \\
       h \ \Phi_{(l,m,s)} &= e^{2\pi i \frac{s}{2}} \ \Phi_{(l,m,s)}
     \end{split}
\end{equation}
Here, $q$ is the $U(1)$ charge of the primary field $\Phi_{(l,m,s)}$,
\begin{equation}
q=\frac{m}{k+2} - \frac{s}{2}
\end{equation}
The corresponding action on the representation labels
is generated by the current  $(0,1,1)$,
which is interpreted
as the spectral flow operator (by $1/2$ units), and
by $(0,0,2)$, which are the representation labels of a
worldsheet supersymmetry generator. The action of these
currents on a representation is inherited by the fusion rules.
Accordingly, they map $(l,m,s)$ to $(l, m+1, s+1)$,
and $(l,m,s)$ to $(l,m,s+2)$.
Using these currents, the fields of the minimal model can be
organized into orbits. The orbits of the spectral flow operator
plays a major role in the discussion of the boundary states.

\subsection{Bulk theory}
To use minimal models for string compactification, we
first have to form tensor products of $r$ minimal models
in such a way that the central charge is equal to $9$.
This is the right central charge for a CFT description
of a Calabi-Yau compactification. In addition, we need
to tensor this with an $SO(2)_1$ current algebra, describing
the uncompactified directions in the light cone gauge.
These products do not give consistent string vacua with
four dimensional $\CN=1$ space time SUSY. But there exists
an orbifold of the tensor product which satisfies all
requirements of a consistent string background.

At this point, let us introduce some notation. First,
we organize the $l$ quantum numbers in a vector
\begin{equation*}
     \lambda = (l_1, \dots, l_r),
\end{equation*}
where $r$ is the number of minimal models. The quantum
numbers $s$ and $m$ are written in an $2r+1$ dimensional
vector:
\begin{equation*}
     \mu = (s_0; m_1, \dots, m_r; s_1, \dots, s_r)
\end{equation*}
One can also define an inner product between these vectors:
\begin{equation}\label{gepprod}
\mu \sbullet \nu = -\frac{\mu_0 \nu_0}{4}
+ \sum_{i=1}^r \left( \frac{\mu_i\nu_i}{2(k_i+2)}
- \frac{\mu_{i+r} \nu_{i+r}}{4} \right)
\end{equation}

Next, we introduce the special $(2r+1)$-dimensional vectors $\beta_0$ 
with all entries equal to 1, and $\beta_j$, $j=1,\ldots,r$, having 
zeroes everywhere  except for the 1st and the $(r+1+j)$th entry which 
are equal to 2. These vectors stand for particular elements in the
symmetry group of the tensor product of the minimal model. 

To implement worldsheet supersymmetry,
one has to  project on representations $(\lambda, \mu)$ 
which satisfy $\beta^{(j)}\sbullet\mu \in \BZ$. To ensure spacetime
supersymmetry, one projects on odd integer $U(1)$ charge,
{\it i.e.}~on representations
$(\lambda, \mu)$ with $2\beta_0 \sbullet \mu \in 2\BZ+1$.

The Gepner model partition function is the partition function
of the tensor product of minimal models, orbifolded by these
symmetries:
\begin{equation}\label{closedpf}
     \frac{(Im\,\tau)^{-2}}{2\vert\eta(q)\vert^2} 
     \sum^{\beta}_{\lambda, (\mu- \overline{\mu}) 
     \in \Lambda}
     (-1)^{s_0} \chi_{\lambda, \mu}(q)
     \chi_{\lambda, \overline{\mu}}(\bar q).
\end{equation}
The symbol $\sum^\beta$ denotes a $\beta$ constrained sum:
The sum is taken  over those $(\lambda, \mu)$ which fulfill
the charge quantization conditions $\mu\sbullet\beta_0 \in 2\BZ+1$ and
$\beta^{(j)} \sbullet \mu \in \BZ$. $\Lambda$ denotes
the lattice spanned by $\beta_0$ and $\beta^{(j)}$.
In the twisted sectors, the right movers are shifted 
with respect to the left movers by
linear combinations of the lattice vectors in $\Lambda$. They
are therefore describing winding modes along $\beta_0$ and
$\beta^{(j)}$.

The symmetries of the Gepner model are given by the subgroup
of symmetries of the tensor product theory which preserve
worldsheet and space time supersymmetry \cite{GrePle}.
These symmetries act as:
\begin{equation}\label{coversym}
     g(\gamma) \Phi_{\lambda\mu; \bl\overline{\mu}} = 
\exp{i\pi\left(\sum_{j=1}^r
     \frac{\gamma_j (\mu_j + \overline{\mu}_j)}{k_j+2} \right)}
     \Phi_{\lambda\mu; \bl\overline{\mu}}  = e^{\pi i 
\beta_{\gamma} \sbullet (\mu
     +\overline{\mu})}
     \Phi_{\lambda\mu; \bl\overline{\mu}},
\end{equation}
where the $\gamma_i$ in $\gamma=(\gamma_1, \dots, \gamma_r)$ specify
the orbifold action in the individual minimal models as in
\eqref{sym}.
$\beta_{\gamma}$ is the vector $(0; 2\gamma;0)$.
For consistency with the projections, we require
$\beta_\gamma \sbullet \beta_0 \in \BZ$.

The new partition function is easy to write down \cite{GrePle}: 
One just has
to include a further vector into the lattice $\Lambda$ and
to project on elements $\mu$  with $\beta_{\gamma}
\sbullet (\mu +\overline{\mu})
\in \BZ$. In lattice language, there are new winding
modes with $\mu-\overline{\mu} = n\beta_\gamma$ coming from twisted sectors.

It is a well-known result by Greene and Plesser that if we
orbifold by all  generators which are compatible with the
charge quantization conditions, we get the mirror theory.
In the case of the quintic, the Greene-Plesser group
is $\BZ_5^3$.

In this paper, we are interested in free orbifolds of a
given Gepner model. Geometrically, we consider free
orbifolds of hypersurfaces in weighted projected space.

Our main example is a free orbifold of the quintic. The
Gepner model for the quintic is given by a tensor product
of five copies of the $k=3$ minimal model with the
appropriate projections. The orbifold
action under consideration is given by
\begin{equation}\label{ourmodel}
\gamma=(0,1,2,3,4) \quad and \quad \beta_1=
\beta_{\gamma} = (0; 0,2,4,6,8; 0, \dots, 0)
\end{equation}
To obtain the mirror of the orbifold applying the Greene-Plesser
construction, we have to orbifold by one further $\BZ_5$
action. This $\BZ_5$ action has to be compatible with
both the GSO projection and the orbifold projection given by
\eqref{ourmodel}. This generator can be chosen as
$\gamma_2=(0,1,3,1,0)$. Taking an orbifold by both $\gamma_1$
and $\gamma_2$ leads to the mirror of the orbifold \cite{AspMor}. 
This can also be understood
in the following way: We first orbifold
the quintic by $\BZ_5^3$, the full Greene-Plesser group of the quintic,
to obtain the mirror of the quintic. Then, we
mod out by a freely-acting $\BZ_5$, which is the quantum
symmetry associated to orbifolding by $\gamma_1$.
In this way the original orbifold is undone. Starting
with the mirror of the quintic and undoing the orbifold by $\gamma_1$
is
the same as starting with the quintic and orbifolding by
$\gamma_1$ and $\gamma_2$.
The full quantum symmetry of the orbifold is given by $\BZ_5\times\BZ_5$,
whose generators are denoted $g$ and $h$. The action of
these symmetries on the D-branes will be discussed
in detail in later sections.

\section{BPS Boundary States for Orbifolds of Gepner Models}\label{sec:GepnerBPS}
The A-type and B-type boundary
states for a single minimal model are  determined using
Cardy's formalism in rational conformal field theory. In particular,
in this framework the boundary states are labeled by the same
labels as the primary fields, we denote boundary states
using capital letters $(L,M,S)$. In the tensor product, 
the labels $L$ are summarized in a vector $\Lambda$ which specifies
the boundary conditions in the WZW part of the minimal models:
$$
\Lambda=(L_1, \dots, L_r)
$$
The $U(1)$ labels are contained in a lattice vector $\Xi$:
$$
\Xi=(S_0; M_1, \dots, M_r; S_1, \dots, S_r).
$$
In this way $(\Lambda, \Xi)$ specifies a consistent boundary
state in the tensor product theory. 

However, as explained for the closed string situation in the
previous section, we have to take an orbifold of the tensor
product theory, to ensure space-time and world-sheet supersymmetry.
The GSO projection consists of a projection on integer charges.
In geometrical language, one would say that the bosonized $U(1)$ in
eq.~\eqref{eq:bozo} is compactified on a circle.

A-type boundary
states have Dirichlet boundary conditions along the circle.
It is well-known from the geometrical context how to construct
A-type branes on a circle by summing over images of the translation
operator \cite{Taylor}. In our context, the translation operator
is given by the spectral flow operator by one half unit. More
explicitly, the projected A-type boundary state can be
written as \cite{RecSch, BruSchI, FuScWa}
\begin{equation}\label{Atype}
|\Lambda, \Xi\rangle_{proj} =
const \sum_{\nu, \nu_i} (-1)^\nu (-1)^{\frac{s_0^2}{2}}
|\lambda, \Xi + \nu \beta_0 + \nu_1\beta^{(1)}
+\dots + \nu_r\beta^{(r)} \rangle.
\end{equation}
Inequivalent A-type boundary
states are  given by orbits of the spectral flow
operator and the operators $\beta^{(r)}$. 
As a consequence, the physically inequivalent choices
for $S_i$ are  encoded in the sum $S=\sum S_i$, which is defined mod $4$.
The $\BZ_2$ operation $S\mapsto S+2$ maps branes
to anti-branes.
The A-type states transform non-trivially
under the symmetry operations \eqref{coversym}.
\begin{equation}\label{abssym}
g(\gamma) \ket{(\Lambda, \Xi)}_{proj} = 
\ket{(\lambda, \Xi+\beta_\gamma)}_{proj}
\end{equation}
Note, however, that because we took an orbit under the spectral
flow the vector $\gamma_0 = (1,1,1,1,1)$ acts trivially,
and consequently the vectors $\gamma$ and $\gamma+\gamma_0$
act in the same way on the boundary state.

B-type boundary states have Neumann boundary conditions on $X$. They
are therefore wrapping the circle given by the bosonized $U(1)$.
In this case, they should be specified by a discrete Wilson line.
In analogy with the situation where D-branes are wrapped
on a geometric  circle (or rather a lattice vector)
this Wilson line is given by
$2\beta_0 \sbullet \Xi$. Similar to the A-type branes, 
the label $S=\sum S_i$  distinguishes between branes
and anti-branes.
The physically inequivalent choices for the  $M_i$ 
can then be encoded in the quantity
$M=\sum w_i M_i$, where $w_i$ are the weights of
the embedding weighted projected space, $w_i=d/(k_i+2)$,
where $d=lcm \{k_i+2 \}$.
Note that this quantity just extracts the $M$
dependent part of the Wilson line $2\beta_0\sbullet \Xi$.
The B-type boundary states are invariant under application
of elements $\beta_{GP}$ of the Greene-Plesser group since per definition
\begin{equation*}
g(\gamma_{GP}) \  \ket{(\Lambda, 2\beta_0\sbullet\Xi)}
= \ket{(\Lambda, 2\beta_0\sbullet(\Xi+\beta_{GP}))} 
= \ket{(\Lambda,2\beta_0\sbullet\Xi)}.
\end{equation*}
This means that of all the generators in \eqref{abssym} 
there remains  only one non-trivially acting symmetry
operation on the B-type states,
\begin{equation}\label{bbssym}
g: M \mapsto M+2.
\end{equation}

Let us now describe boundary states on the orbifold.
A set of A-type boundary states can be obtained by simply
projecting the boundary states in \eqref{Atype} by the
additional orbifold operation. In formulas:
\begin{equation}
|\Lambda, \Xi\rangle_{orb} =
const \sum_{\nu, \nu', \nu_i} (-1)^\nu (-1)^{\frac{s_0^2}{2}}
|\lambda, \Xi + \nu \beta_0 +\nu' \beta_{\gamma} + \nu_1\beta^{(1)} +\dots +
\nu_r\beta^{(r)} \rangle.
\end{equation}
This set of A-type states is given by orbits of the spectral flow
operator \emph{and} the additional orbifold operator.
To obtain the A-type states for the orbifold of the quintic,
plug in the explicit generator $\gamma_1$ in \eqref{ourmodel}.

Let us turn to the B-type states. There are more B-type
states on the orbifold  than on the original model.
The D-branes of the original
model split up into several fractional branes.
The reason for this is that the closed string orbifold partition
function has additional twisted sector states, the lattice $\Lambda$
in \eqref{closedpf} is enhanced by an additional vector $\beta_{\gamma}$,
which can be wrapped. There are two labels needed to
distinguish between different B-type branes,
the label $M$ and $M^1= d\sum \frac{M_i \gamma_i}{k_i+2}$.
Explicit formulas for the B-type states are provided in the
appendix.

It is instructive to look at the situation in the language of
\cite{DiaDou}. They argue that B-type boundary states (with
$\Lambda = (0,\dots, 0)$) can be thought of as the restriction
of the fractional brane states of a $\BC^r/\BZ_K$ orbifold describing
the Landau Ginzburg phase of the linear $\sigma$-model 
to the Calabi-Yau hypersurface. In this picture, the B-type
states on a $\BZ_N$ orbifold correspond to fractional branes
on $\BC^r/\BZ_K \times \BZ_N$, which are characterized
by the irreducible representations of the  orbifold group.
For $\BZ_N$ groups, these representation labels are just given by
a phase, and this phase can be mapped to the Wilson-line label
introduced before.

The states on the orbifold are only invariant under the Greene-Plesser group
of the orbifold, which is a subgroup of the Greene-Plesser group
of the original model. The quantum symmetry of the orbifold model
is bigger than that of the covering space, and the different fractional
branes can be related by applying the symmetry generators.
For the quintic orbifold, the B-type states are labeled by $M=\sum M_i$ and
$M^1= M_2 + 2M_3 +3 M_4 +4 M_5$. They transform non-trivially under
the $\BZ_5 \times \BZ_5$ quantum symmetry:
\begin{equation}\label{BPSq}
     \begin{split}
       g: M &\to M+2, \quad M^1 \to M^1 \\
       h: M &\to M, \quad M^1 \to M^1 +2
     \end{split}
\end{equation}
$g$ and $h$ can be written in matrix form if we
arrange the BPS boundary states in
a 25-dimensional vector with entries $(M, M^1)$.
\begin{equation*}
     ( (0,0), (2,0), (4,0), (6,0), (8,0), (2,0), (2,2), \dots (8,8) )^t
\end{equation*}
On these vectors, $g$ acts as a $25\times 25$ matrix, consisting
of 5 dimensional shift matrices on the diagonal. On the other hand,
$h$ acts as a shift matrix on those $5\times 5$ blocks.

These symmetry operations specify the transformation properties
of the BPS branes (periods about the Gepner point) under
the quantum symmetry, and can  be compared with
the results obtained from K-theory considerations.
The generator $g$ corresponds to the additional $\BZ_5$
symmetry  at the Gepner point. Geometrically,
its action on branes is given by the monodromy matrix
around the Gepner point. The group generated by $h$ is
a universal symmetry group and should be compared to
the quantum symmetry discussed in \S\ref{sec:QuantSym}. 
In \S\ref{KofX}  these symmetry operations are worked out
for the example at hand.

So far, we only considered orbifold groups generated by
a single group element. Of course, it is possible to repeat the same
procedure for an additional generator. In particular, 
the full Greene-Plesser group can be divided out. In the
bulk, this yields the theory on the mirror and the procedure
outlined above provides us with a method to confirm that
mirror symmetry extends to the open string sector
\cite{Vafa, Kont, StYaZa}.
A-type boundary states on the mirror obtained through an orbifold
by the Greene-Plesser group are organized in terms of orbits
of this  group. The resulting open string sector inherits
that structure. Therefore, the projected (A-type) partition function
coincides with that found in \cite{RecSch} for B-type states
on the original manifold \cite{BruSchI}. 
The Wilson line $M$ on the  B-type side
becomes an orbit label on the A-type side.
Conversely,
we can also understand A-type states as B-types on
the mirror: Taking successive orbifolds, the B-type
states split up further into fractional branes
and a new label $d\sum \gamma_i M_i/(k_i+2)$
has to be introduced after each step. After all Greene-Plesser
generators are modded out, the states will be orbits of
only the spectral flow operator. In this way, 
A-type boundary states can be obtained as B-type states on the mirror.

This interpretation provides a convenient way to specify
the A-type branes in terms of $d\sum \gamma_i M_i/(k_i+2)$ for
all Greene-Plesser operators. For the quintic we will
use the following labels for A-type branes:
$M=\sum M_i$, $M^1 = M_2 + 2M_3 +3 M_4 +4 M_5$
and $M^2 = M_2 +3M_3 + M_4$. These states transform
non-trivially under $\BZ_5\times\BZ_5\times\BZ_5$,
which is the quantum symmetry of the mirror of the orbifold.
\begin{equation}    \label{BPSqa}
     \begin{split}
     G: M &\to M+2, \quad M^1 \to M^1 \quad M^2\to M_2\\
     H: M &\to M, \quad M^1 \to M^1 +2 \quad M_2 \to M_2 \\
     H': M &\to M, \quad M^1 \to M^1  \quad M_2 \to M_2+2
     \end{split}
\end{equation}
In \S\ref{KofX}, this symmetry action will be compared with
the results obtained from K-theory.

\subsection{The intersection form for the orbifold models}

In this section, we will compute the intersection
numbers for the BPS states on the orbifold. Since
the torsion state we are looking for is supposed to
have zero intersection with all branes, this computation
gives a first hint on the symmetry properties of
the torsion brane.

In boundary conformal field theory, the intersection
form of two branes can be computed as the Witten index
in the open string sector, $\tr_R (-1)^F$.
To compute the intersection numbers on the orbifold of the
quintic $(k=3)^5$,
we  use    the formulas and notation  of
\cite{BDLR}, where the expression for the A-type
boundary states with $\Lambda=(0,0,0,0,0)$ was given as
\begin{equation}
I^A= \prod_{i=1}^5 (1-g_i^{-1}).
\end{equation}
Here, the $g_i$ are the symmetry operations in \eqref{abssym},
with $g_1=g((1,0,0,0,0)), \dots, g_r=((0,0,0,0,1))$. 
Because of the GSO projection
the generators $g_i$
are not independent and we can express one of them, e.g. $g_5$, in terms
of the other generators, $(g_5)^{-1}= g_1 g_2 g_3 g_4$.
This leads to the following expression for the intersection
form \cite{BDLR} :
\begin{equation*}
    I^A=  (1-g_1^4)(1-g_2^4)(1-g_3^4)(1-g_4^4)(1-g_1g_2g_3g_4).
\end{equation*}
For the B-type states on the
quintic, all the $g_i$ get identified, leaving only the
non-trivial generator $g$ of \eqref{bbssym}.
The result for the
intersection matrix is therefore:
\begin{equation}\label{eq:QuinticBint}
I^B= (1-g^{-1})^5,
\end{equation}
where $g$ is the generator mapping $M\to M+2$, where $M=\sum M_i$.

On the orbifold we are in an intermediate situation, where
some of the $g_i$ get identified. The intersection forms can
be written in terms of the quantum symmetry generators $g,h$ for the
B-type states and in terms of $G,H,H'$ for the A-type states.
Their action on the boundary states is given in
\eqref{BPSq} and \eqref{BPSqa}.

Using $g$ and $h$, the intersection form for B-type states
on the orbifold is given by
\begin{multline}\label{eq:XBint}
I^B_{orb} =g^4 ( -h^4 -h^3 -h^2-h -1) + g^3 (2h^4 +2h^3 +2h^2 +2h +2) \\
+ g^2( -2h^4 -2h^3 -2h^2 -2h-2) +g(h^4 + h^3 +h^2 +h+1)
\end{multline}
This means that the intersection matrix depends only on $g$,
not on $h$. The intersection matrix between two states which
differ only in $M^1$, not in $M$ is zero, and that of two
states which differ only in $M$ equals the intersection
number on the quintic. For two states differing by a general
application of $g^nh^m$, the intersection number is independent
of $m$.

Using the generators in \eqref{BPSqa}, the
intersection form for the A-type states can  be written as
\begin{multline}
I_{orb}^A = G^4 ( -1 - H^4 H^{'4} - H^3 H^{'4} - H^2 - H H^{'2}) \\
+ G^3 ( H^4 H^{'4} + H^3 H^{'4} + H^2 H^{'3} + H^2
+ H H^{'4} + H^{'4} + H H^{'2} + H^{'} + H^4 H' + H^3 H^{'2}) \\
+G^2 (- H^2 H^{'3} - H H^{'4} - H^{'4} - H^4 H^{'3}
- H' - H^4 H' - H^3 - H^3 H^{'2} - H^2 H' - HH') \\
+ G ( H^4 H^{'3} + H^3 + H^2 H' + HH' +1)
\end{multline}
States, which only differ by an application of
$H$, have zero intersection number. The intersection
number is determined by the group actions $G$ and $H'$.

\section{Non-BPS States on the Gepner Orbifold}\label{sec:GepnerNonBPS}
\begin{floatingfigure}{2.75in}
   \mbox{\includegraphics[width=2.5in]{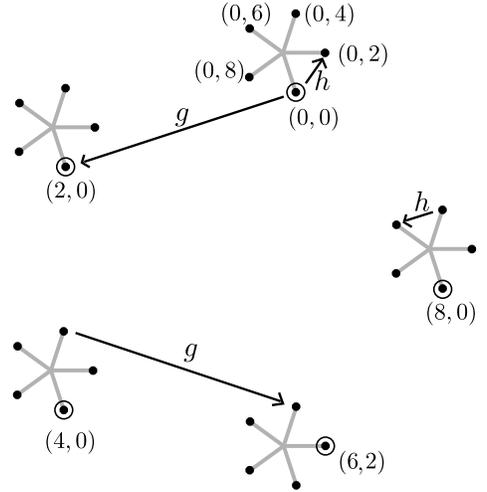}}
   \caption{Graphical depiction of a torsion state: Each dot
      represents an $L=0$ BPS brane. The symmetry generator $h$ acts
      within each small pentagon; $g$ maps the branes
      of one pentagon onto the next. The 5 branes marked by circles
      can decay to a stable torsion brane.}\label{fig:quiver}
\end{floatingfigure} 

The D-branes we are interested in are non-BPS but stable.
Their stability is guaranteed because, though they are not
charged under any RR field, they carry a discrete torsion
charge. We are going to construct the torsion brane as a
bound state of the BPS branes, whose boundary state description
has been given in the previous section. The idea is to form a bound
state such that all the BPS charges cancel, leaving only a net torsion
charge.

Let us first consider the B-type states. Since the intersection
form does not depend on $h$, we conclude that the application
of $h$ can only shift the torsion charge.
Starting from a given brane $\ket{(\Lambda,\Xi)}$, the RR
charge can be canceled by adding a full orbit of
$g$-images to that brane. However, for these branes there
is no obstruction to decay to the vacuum. To be left
with a non-trivial torsion charge, we modify the $g$-action
on the orbit by suitable powers of $h$. This is possible,
since the state $h^m \, g \ket{(\Lambda,\Xi)}$ has the same
intersections as $g \ket{(\Lambda,\Xi)}$, but differs from
it by an action on $M^1$.
Summarizing, we add up the following five branes (taking into
account the appropriate Chan-Paton indices):
\begin{multline}\label{tor}
|(M,M^1)\rangle + h^{n_1} g\ket{(M,M^1)} + h^{n_2} g^2
\ket{(M,M^1)}\\ +
h^{n_3} g^3 \ket{(M,M^1)}+ h^{n_4} g^4 \ket{(M,M^1)} \\
= \sum_m h^{n_m} g^m \ket{(M,M^1)},
\end{multline}
where
\begin{equation*}
     \sum n_m \neq 0 \mod 5.
\end{equation*}
Note, that there are tachyons propagating
between the individual branes of this stack of branes. Therefore,
this unstable stack of branes can decay into a stable single particle state.
We argued that this state is the torsion
brane we are looking for. This claim will get further support
from the K-theory analysis in the following section,
where  the K-theory classes of all BPS
boundary states and their symmetry properties will be identified. 
Taking
a sum of K-theory classes as in \eqref{tor} will result in a
torsion class.

The A-type states can be discussed similarly. From their intersection
form it can be concluded that to cancel the RR charge a sum over all
powers of $G$ and $H'$ has to be taken. To have a remaining
torsion charge, $M^1$ should not cancel out. This means that
we consider the following stack of branes:
\begin{equation}
\sum_{m, m'} H^{n_{(m,m')}} G^m (h')^{m'}\ket{(M,M^1,M^2)},
\end{equation}
where
$$
\sum n_{(m,m')} \neq 0 \mod 5.
$$
Also for the A-type branes this picture can be verified
using K-theory arguments.

To summarize, we have shown that for the free orbifold of the quintic
generated by $\gamma_1$ there are torsion B-type and A-type
states. The A-type states can be reinterpreted as B-type
states on the orbifold by $\gamma_1$ and $\gamma_2$. One
might ask what happens if we take an orbifold by the third
Greene-Plesser generator. Since this generator does not
preserve the projection by $\gamma_1$, the torsion states
get projected out. This is what is expected from geometry,
since the mirror of the quintic has no torsion in K-theory.

\section{The K-theory of $X$}\label{sec:ExampleX}
\subsection{The K-theory of the Quintic}
First let us review the K-theory of the quintic, $Y$.

The even dimensional
cohomology of $Y$ has generators $1,\xi_2,\xi_4$ and $\xi_6$, with relations
$\xi_2^2=5\xi_4$, $\xi_2\cup\xi_4=\xi_6$ and $\xi_4^2=0$. The total Chern class
of $Y$ 
is
\begin{equation*}
    c(Y)=1+50\xi_{4}-200\xi_{6}
\end{equation*}
$\coho{3}{Y}$ has 204
generators $\zeta_i$, and we have $\zeta_i\cup\zeta_j=\omega_{ij}\xi_6$ for
some constant antisymmetric matrix $\omega_{ij}$. Also
$\xi_2\cup\zeta_i=\xi_4\cup\zeta_i=0$. 

Let $H$ be the hyperplane line bundle on $Y$, and $D$ be the corresponding
hyperplane divisor. Let $C$ be a degree-one rational curve (a $\BP^1$)
in $Y$ and $p$ a point in $Y$. $K^0(Y)$
has the following generators:
\begin{center}
\begin{tabular}{|c|c|c|c|c|c|}
\hline
&$r$&$c_1$&$c_2$&$c_3$\\
\hline\hline
$\CO$&$1$&$0$  & $0$ & $0$\\
$a=i_!\CO_D=H-\CO$&$0$&
$\xi_2$  & $0$ & $0$\\
$b=i_!\CO_C$&$0$&$0$  & $-\xi_4$ & $2\xi_6$\\
$c=i_!\CO_p$&$0$&
$0$  & $0$ & $2\xi_6$ \\
\hline
\end{tabular}
\end{center}
These correspond simply to a 6-brane wrapped on $Y$, a 4-brane wrapped on $D$,
a 2-brane wrapped on $C$  and a 0-brane sitting at the point $p$.

The relations in the K-theory ring are similar to what we had in cohomology:
$a^2=5b$, $ab=c$ and $b^2=0$. For later use, it is also worthwhile 
knowing that complex conjugation acts by: $\overline{a}=-a+5b-5c$, 
$\overline{b}=b-2c$ and $\overline{c}=-c$.

$K^1(Y)$ has generators $u_i\in \tilde K^0(Y\times S^1)$ whose only
nonvanishing chern class is $c_2(u_i)=-\zeta_i\cup\phi$, where $\phi$ is the
fundamental class of $S^1$.

The intersection form on $K^0(Y)$ is
\begin{equation}
    (v,w)=Ind(\overline{\partial}_{v\otimes\overline{w}})
      =\int_Y ch(v\otimes\overline{w})Td(Y)
\end{equation}
In the above basis, this is the matrix
\begin{equation}
    \Omega=
    \begin{pmatrix}
    0&-5&-1&-1\\
    5& 0& 1& 0\\
    1&-1& 0& 0\\
    1& 0& 0& 0
    \end{pmatrix}
\end{equation}
The monodromies are generated by the following operations:
$M_{r=\infty}: v\mapsto v\otimes H$ at large radius,
$M_c: v\mapsto v-Ind(\overline{\partial}_v)\CO$ at the conifold
and $M_g= M_c M_{r=\infty}$ at the Gepner point.
With respect to the above basis, these are represented by the matrices
\begin{equation}
    M_{r=\infty}=
    \begin{pmatrix}
    1&0&0&0\\
    1&1&0&0\\
    0&5&1&0\\
    0&0&1&1
    \end{pmatrix},\qquad
    M_{c}=
    \begin{pmatrix}
    1&-5&-1&-1\\
    0&1&0&0\\
    0&0&1&0\\
    0&0&0&1
    \end{pmatrix},\qquad
    M_{g}=
    \begin{pmatrix}
    -4&-10&-2&-1\\
    1&1&0&0\\
    0&5&1&0\\
    0&0&1&1
    \end{pmatrix}
\end{equation}
Note that $M_g^{5}=1$. 

To make the connection to CFT at the Gepner point, we need to find a set of five
K-theory elements which are cyclically permuted under the $\BZ_5$ action of
$M_g$. Set
\begin{equation}
   V_{n}= M_g^{-n} \CO
\end{equation}
or, explicitly,
\begin{equation}
\begin{split}
V_{1}&=\CO-a+5b-5c\\
V_{2}&=-4\CO+3a-10b+5c\\
V_{3}&=6\CO-3a+5b\\
V_{4}&=-4\CO+a\\
V_{5}&=\CO
\end{split}
\end{equation}
Together, these span an index-25 sublattice of $K^0(Y)$ -- only 2-brane and
0-brane charges which are multiples of 5 are realizable as linear combinations
of the $V_n$. The intersection pairing between these generators is given by
the $5\times5$ matrix
\begin{equation*}
  \begin{pmatrix}
     0&  5&-10& 10& -5\\
    -5&  0&  5&-10& 10\\
    10& -5&  0&  5&-10\\
   -10& 10& -5&  0&  5\\
     5&-10& 10& -5&  0
  \end{pmatrix}
\end{equation*}
which agrees with the intersection form \eqref{eq:QuinticBint} for the
B-type boundary states  at the
Gepner point. We therefore identify the B-type boundary states constructed
in the CFT as BPS
representatives of the above K-theory classes.

In fact, as noted by Diaconescu and Douglas \cite{DiaDou}, the sublattice of $K^0(Y)$
which arises in this way is precisely the pullback of the K-theory of
$\BC P^4$ under the embedding $Y\hookrightarrow \BC P^4$. Their rationale
for this was that, for the purpose of studying the B-type states, one can
ignore the superpotential in the N=2 gauged linear $\sigma$-model.

If one ignores the superpotential, then in the large-radius phase of the
GL$\sigma$M, one has a sigma model with target space $\tilde Y=\CO_{\BP^4}(-5)$,
the total space of the canonical bundle of $\BP^4$. This is a noncompact
Calabi-Yau manifold, and its K-theory, $K(\CO_{\BP^4}(-5))=K(\BP^4)$.

\subsection{The K-theory of $X$}\label{KofX}
Now we turn to the same computation on $X=Y/\BZ_5$, where we mod out by
\eqref{eq:FirstZ5}. The cohomology ring of
$X$ is a little simpler than for the quintic. $\coho{ev}{X}$ is generated by
$1,\xi'_2,\chi,\xi'_2{}^2,$ and $\xi'_2{}^3$, with relations
$$ \chi\cup\chi=5\chi=\chi\cup\xi'_2=\xi'_2{}^4=0$$
The total Chern class of $X$ is 
\begin{equation*}
    c(X)=1+10\xi'_{2}{}^{2}-40\xi'_{2}{}^{3}
\end{equation*}
Under the projection $\pi_{1}:Y\to X$, we have 
$\pi_{1}^{*}(\xi'_{2})=\xi_{2}$, and, consequently, 
$\pi_{1}^{*}(c(X))=c(Y)$.

$\coho{3}{X}$ has generators $\zeta_i$, $i=1,\dots 44$, and $\coho{5}{X}$ has
a single generator $\chi\cup\zeta$ (the cup product being independent of $i$).

Now we turn to the $K$-theory. As before let $H$ be the hyperplane line
bundle and let $L$ be the flat line bundle
\begin{equation*}
    L= (Y\times \BC)/\BZ_5
\end{equation*}
where $\BZ_5$ act by multiplication by a fifth root of unity on $\BC$. A
basis for $K^0(X)$ is
\begin{center}
\begin{tabular}{|c|c|c|c|c|c|}
\hline
&$r$&$c_1$&$c_2$&$c_3$\\
\hline\hline
$\CO$&$1$&$0$  & $0$ & $0$\\
$a=H-\CO$&$0$&
$\xi_2$  & $0$ & $0$\\
$\alpha=L-\CO$&0&$\chi$&$0$&$0$\\
$a^2$&$0$&$0$  & $-\xi_2^2$ & $2\xi_2^3$\\
$a^3$&$0$&
$0$  & $0$ & $2\xi_2^3$ \\
\hline
\end{tabular}
\end{center}
where $a\otimes\alpha=0$. Complex conjugation acts by 
$\overline{a}=-a+a^{2}-a^{3}$ and $\overline{\alpha}=-\alpha$.

Note that there is no preferred choice for $a$.
Shifting $a\to a'=a+\alpha$ yields exactly the same ring. For flat line bundles,
there is a preferred one which is actually trivial. There is no similar preferred
choice if the line bundle isn't flat.

$K^1(X)\subset \tilde K^0(X\times S^1)$ is given by generators $u_i$ whose
nonvanishing Chern classes are $c_2(u_i)=-\zeta_i\cup\phi$ and $\beta$, whose
only nonvanishing Chern class is $c_3(\beta)=2\chi\cup\zeta\cup\phi$.

The intersection form on $K^0(X)$ is given with respect to the above 
basis (omitting $\alpha$, which has zero intersection with everything) by
\begin{equation}
    \Omega=
    \begin{pmatrix}
    0&-1&-1&-1\\
    1& 0& 1& 0\\
    1&-1& 0& 0\\
    1& 0& 0& 0
    \end{pmatrix}
\end{equation}
and the monodromies are generated by the following operations:
$M_{r=\infty}: v\mapsto v\otimes H$ at large radius,
$M_c: v\mapsto v-5Ind(\overline{\partial}_v)\CO$ at the conifold and
$M_g= M_c M_{r=\infty}$ at the Gepner point.

The conifold monodromy is easy to understand by the considerations of
\S\ref{sec:Pairings}. At the conifold point, 5 mutually local particles
become massless. They are the wrapped D6-brane and the D6-brane with $p$
torsion D4-branes on it, for $p=1,2,3,4$.
These lie in the K-theory classes $\CO, L, L^2,L^3,L^4$. By
\eqref{eq:GeneralConifoldMonodromy}, the monodromy is
\begin{equation}\label{eq:sumL}
    \begin{split}
    v\mapsto& v- 
    (v,\CO)\CO-(v,L)L-(v,L^{2})L^{2}-(v,L^{3})L^{3}-(v,L^{4})L^{4}\\
    &=v-(v,\CO)(\CO+L+L^{2}+L^{3}+L^{3}+L^{4})\\
    &=v-5(v,\CO)\CO
    \end{split}
\end{equation}
where we used the fact that the intersection
pairing with $L^p$ is independent of $p$ and $\CO+L+L^2+L^3+L^4=5\CO$. 
So, for the purposes of the considerations of \S\ref{sec:Pairings}, we 
just have five particles carrying 6-brane charge becoming massless at the 
conifold (instead of one such particle for the quintic).

With respect to the above basis,
the above monodromies are represented by the matrices
\begin{equation}
    M_{r=\infty}=
    \begin{pmatrix}
    1&0&0&0\\
    1&1&0&0\\
    0&1&1&0\\
    0&0&1&1
    \end{pmatrix},\quad
    M_{c}=
    \begin{pmatrix}
    1&-5&-5&-5\\
    0&1&0&0\\
    0&0&1&0\\
    0&0&0&1
    \end{pmatrix},\quad
    M_{g}=
    \begin{pmatrix}
    -4&-10&-10&-5\\
    1&1&0&0\\
    0&1&1&0\\
    0&0&1&1
    \end{pmatrix}
\end{equation}

Another way to think about $\beta$ is as the push-forward $i_!\CO_\ell$ of
the trivial line bundle on the torsion 1-cycle, $\ell$. In that case, the
torsion pairing between
$\alpha$ and $\beta$ consists of computing the holonomy of $L=\CO+\alpha$
restricted to $\ell$.

At a generic point in the moduli space, the quantum symmetry is $\BZ_5$, generated
by $v\mapsto v\otimes L$, {\it i.e.}~$\CO\to \CO+\alpha$
and $u_i\mapsto u_i+\beta$, with all other basis vectors held fixed.
Note that the monodromies (in particular, the conifold monodromy) commute with this action.
At the Gepner point, the quantum symmetry is enhanced to $\BZ_5\times\BZ_5$,
where the second $\BZ_5$ is represented by $M_g$ above acting on $K^0(X)$, and acts 
trivially on $K^{1}(X)$.
These symmetries preserve both the intersection pairings on $K^0(X)$ and
$K^1(X)$ and the torsion pairing: $K^0(X)_{tor}\times K^1(X)_{tor}\to \BZ/5\BZ$.

To make contact with the B-type states in the conformal field theory, we
follow the procedure above. We write down the orbit of $\CO$ under the
$\BZ_5\times\BZ_5$ quantum symmetry. The orbit generated by $M_g$ is
\begin{equation}
\begin{split}
V_{1,5}&=\CO-a+a^2-a^3\\
V_{2,5}&=-4\CO+3a-2a^2+a^3\\
V_{3,5}&=6\CO-3a+a^2\\
V_{4,5}&=-4\CO+a\\
V_{5,5}&=\CO
\end{split}
\end{equation}
The generator of the other $\BZ_5$ shifts $\CO\to \CO+\alpha$.
Noting that $6\alpha=-4\alpha=\alpha$, this means that it shifts
$V_{n,5}\to V_{n,5}+\alpha$. So full $\BZ_5\times\BZ_5$ orbit consists
of 25 K-theory classes
\begin{equation}
V_{n,m}=V_{n,5}-m\alpha
\end{equation}
Their intersection form is clearly independent of $m$, and the
$n$-dependence is given by the $5\times5$ matrix
\begin{equation*}
  \begin{pmatrix}
     0& 1&-2& 2&-1\\
    -1& 0& 1&-2& 2\\
     2&-1& 0& 1&-2\\
    -2& 2&-1& 0& 1\\
     1&-2& 2&-1& 0
  \end{pmatrix}
\end{equation*}
in agreement with the CFT result \eqref{eq:XBint}.
So we identify the 25 B-type boundary states as BPS representatives of the
25 K-theory classes $V_{n,m}$.

Having made this identification, we can give a further,
{\it a posteriori}
argument for the correctness of \eqref{eq:sumL}, in particular, for our
assertion that $M_c$ acts trivially on the torsion subgroup. Using this
identification, the results of
\S\ref{sec:GepnerBPS} show that $M_g$ acts trivially on the torsion subgroup.
We already had that $M_{r=\infty}$ acts trivially on the torsion subgroup, so
we conclude that $M_c$ acts
trivially on the torsion subgroup, as we determined above.

To form a torsion brane representing the class $\alpha$, we sum
\begin{equation*}
 V_{1,n_1}+V_{2,n_2}+V_{3,n_3}+V_{4,n_4}+V_{5,n_5}
\end{equation*}
with $\sum n_i=1\mod 5$. This cancels the BPS charges, leaving just the
torsion charge. After tachyon condensation, what remains is a non-BPS 
brane carrying torsion charge $\alpha$.

We have constructed torsion D-branes at large radius and at the 
Gepner point. This likely means that there is a region of the moduli 
space, containing both large-radius and the Gepner point, in which 
the torsion branes are stable. It is not clear that they are stable 
everywhere. Indeed, in a related situation, Gopakumar and Vafa 
\cite{GopVaf} have argued that the torsion branes might be unstable 
near the conifold point.

The situation is this: at the conifold, the D6-brane (with or without 
torsion charge) is becoming massless. If the mass of the ``pure" 
torsion brane stays finite at conifold point, then it clearly becomes 
unstable to decaying into a D6-brane,anti-D6-brane pair (carrying net 
torsion charge). If this is the case, then there is a curve of 
marginal stability surrounding the conifold point, across which the 
torsion 4-brane becomes unstable.

Of course, the above scenario depends on the behaviour of the mass of 
the torsion brane as we approach the conifold. Since we don't really 
know how it behaves, we cannot make a definitive prediction.

\section{The Permutation Orbifold}\label{sec:W}
Now we consider the Calabi Yau manifold $W=X/\BZ_5$, where we mod
out by the freely acting $\BZ_5$, \eqref{eq:SecondZ5}. 

\subsection{Action of the permutation group}\label{sec:Wconf}
In the Gepner model description of $W$ there is an additional
$\BZ_5$ action, which has to be implemented on the boundary
states. The action of that $\BZ_5$ on the primary fields is
given by:
\begin{equation}
\begin{split}
\sigma: (l_1,l_2,l_3,l_4,l_5) &\mapsto (l_5,l_1,l_2,l_3,l_4), \\
\sigma: (s_0; m_1,m_2,m_3,m_4,m_5; s_1,s_2,s_3,s_4,s_5)
&\mapsto
(s_0; m_5,m_1,m_2,m_3,m_4; s_5,s_1,s_2,s_3,s_4).
\end{split}
\end{equation}
It is easy to see how this action translates into an action
on boundary states in the tensor product theory:
\begin{equation}
\begin{split}
\Lambda = (L_1,L_2,L_3,L_4,L_5) &\mapsto (L_5,L_1,L_2,L_3,L_4), \\
(S_0; M_1,M_2,M_3,M_4,M_5; S_1,S_2,S_3,S_4,S_5) &\mapsto
(S_0; M_5, M_1,M_2,M_3,M_4; S_5,S_1,S_2,S_3,S_4)
\end{split}
\end{equation}
In the following, it is assumed that the boundary states under
consideration carry the same $L$ labels in all minimal models
$\Lambda=(L,L,L,L,L)$.
In particular, we are interested in the B-type boundary states
with $\Lambda=(0,0,0,0,0)$. As a first step, let us consider
an action of the permutation group in the unorbifolded $(k=3)^5$
theory. The B-type
branes, being labeled by $M=\sum M_i$, are invariant under the
action of the permutation group. This is no longer the case
on the orbifold $X$, since $\gamma$ acts differently on the
individual minimal model factors. The permutation acts on the
branes on the orbifold as:
\begin{equation}
|(0,0,0,0,0), M, M^1\rangle \mapsto |(0,0,0,0,0), M, (M^1+M)\rangle
\end{equation}
As a consequence, the five states with $M=0$ are invariant under
the cyclic permutation. The other 20 states can be organized
in four $\BZ_5$ orbits of length five. Note that the action
of the permutation symmetry is the same as that of an
($M$ dependent) power of $h$, meaning that the four orbits
are formed in such a way that the resulting brane configuration can
be lifted to the quintic. In particular, each orbit is labeled
by $M$. 

To study the branes on the orbifold W it is important
to realize that the generator $\sigma$ of the cyclic permutation
commutes with the generator $g(\gamma_1)$ only up to a generator
of the GSO projection $g_0$. This reflects the fact that
geometrically these two actions only commute up to
projective equivalence of the embedding projective space.
The full orbifold group is
therefore a non-abelian group $\Gamma$ with the relation:
\begin{equation}
g(\gamma_1) \sigma = g_0 \sigma g(\gamma_1)
\end{equation}
The commutator subgroup is just the $\BZ_5$ subgroup generated by $g_0$, and
we have the exact sequence of groups
\begin{equation}
   0\to\BZ_5\to \Gamma\to\BZ_5\times \BZ_5\to 0
\end{equation}
The quantum symmetry group of the orbifold model is
$\Gamma/[\Gamma,\Gamma]=\BZ_5\times\BZ_5$.

 To classify the branes, we have to find all irreducible
representation of $\Gamma$. Each representation can be
used as an action on the Chan-Paton factors and therefore
specifies a brane. We immediately see that the orbits
of four which we found above correspond to irreducible
representations of the orbifold group, where $g_0$ acts
as a phase multiplication:
\begin{equation}
g(\gamma_1) \sigma = e^{\frac{\pi i}{5} M} \sigma g(\gamma_1) \quad \quad
M=0,2,4,6,8
\end{equation}
In addition to this we find $25$ one dimensional representations,
where $g_0$ acts trivially. In that case, $g(\gamma_1)$ and
$\sigma$ generate an abelian group. 
To check if this covers all irreducible inequivalent representations,
we apply a lemma of Burnside, which states that the sum of
the square of the dimensions of the irreducible representations
equals the order of the group. Since the order of the group is
$5^3$, these $29$ representations are indeed all irreducible
inequivalent representations. 
The representations of the full non-abelian group can be
reinterpreted in terms of representations of the group
generated by $g(\gamma_1)$ and $\sigma$. This is the group
which has a direct interpretation in geometry. We see
that for $M=0$ this group acts in one-dimensional representations,
and we have the usual picture of fractional branes labeled
by two phases. For $M\neq 0 $ this groups acts in a
projective representation on the Chan-Paton factors.
In other words, we have turned on discrete torsion. The two-cocycle
in $H^2(\Gamma, U(1))$ 
is specified by the phase factor appearing in the
projective representation:
\begin{equation}\label{eq:DiscTor}
\epsilon(g(\gamma_1)^m \sigma^n, g(\gamma_1)^p \sigma^q))
= e^{\frac{\pi i}{5}M(mq-pn)}
\end{equation}

The quantum symmetry of the model,
$\Gamma/[\Gamma,\Gamma]=\BZ_5\times\BZ_5$, is the \emph{same} quantum
symmetry group that we see at a generic point in the moduli space. In contrast
to the previous cases, there is no enhanced symmetry at the Gepner point.

\subsection{The K-theory of $W$}
$\coho{ev}{W}$ is
generated by $1,\xi''_2,\chi_1,\chi_2,\xi''_4$ and $\xi''_6$, with relations
\begin{equation*}
\xi''_2{}^2=25\xi''_4,\qquad \xi''_2\cup\xi''_4=\xi''_6,\qquad
\xi''_2\cup\chi_i=\xi''_4\cup\chi_i=5\chi_i=\chi_i\cup\chi_j=0
\end{equation*}
The total Chern class of $W$ is 
\begin{equation}
   c(W)=1+10\xi_4''-8\xi''_6
\end{equation}
Under the projection $\pi_2:X\to W$,
\begin{equation*}
   \pi_2^*(\xi''_2)=5\xi'_2,\qquad\pi_2^*(\xi''_4)=\xi'_{2}{}^{2},
   \qquad \pi_2^*(\chi_1)=\chi,\qquad \pi_2^*(\chi_2)=0
\end{equation*}
and consequently $\pi_2^*(c(W))=c(X)$.

Note that the generator of $\coho{2}{W}$ pulls back to 5 times the generator of
$\coho{2}{X}$. Equivalently, the generator of $\homo{2}{X}$ pushes forward to 5 times
the generator of $\homo{2}{W}$. Recall that the Cartan-Leray spectral sequence
tells us that $\pi_{1*}: \homo{2}{Y}\to \homo{2}{X}$ is an isomorphism, since
$\homo{2}{\BZ_5}=0$. On the other hand, applying the Cartan-Leray spectral
sequence to $W=Y/(\BZ_5\times\BZ_5)$, one has an extension
\begin{equation}
   0\to \homo{2}{Y}\xrightarrow{\pi_*} \homo{2}{X}\to \BZ_5\to 0
\end{equation}
since $\homo{2}{\BZ_5\times\BZ_5}=\BZ_5$. There are two possible extensions:
either the sequence splits, or it does not. In fact, Aspinwall and Morrison
\cite{AspMor} show that the generator
of $\homo{2}{W}$ is represented by an elliptic curve, $E$, which is not the
image of an element in $\homo{2}{Y}$. Hence the sequence does not split,
$\pi_*=\pi_{2*}\circ\pi_{1*}$ is multiplication by 5 and therefore so
is $\pi_{2*}$.

Note that we are in the situation discussed in
\S\ref{sec:QuantSym}, where $\coho{3}{W}$ is torsion-free, whereas
$\homo{2}{G}\neq0$. Viewed as an orbifold CFT, the model on $W$ admits discrete
torsion, but this can be continuously turned off.
As a consequence, the K\"ahler moduli space of $W$ is a fivefold cover of the
moduli  space of $X$, branched at the Gepner point and the large-radius point.
As we saw in \S\ref{sec:Wconf}, there is no enhancement of the quantum symmetry
at the Gepner point and so the Gepner point is now a smooth point in the moduli
space. Since we have a fivefold cover, the coordinate on the moduli space is now
$\psi$ rather than $\psi^5$. Points related by multiplying $\psi$ by
a fifth root of unity are related by adding one unit of discrete torsion
\eqref{eq:DiscTor}. Hence they are physically inequivalent, as was already
noted by \cite{AspMor}.

A basis for $K^0(W)$ is
\begin{center}
\begin{tabular}{|c|c|c|c|c|c|}
\hline
&$r$&$c_1$&$c_2$&$c_3$\\
\hline\hline
$\CO$&$1$&$0$  & $0$ & $0$\\
$a=S-\CO$&$0$&
$\xi''_2$  & $0$ & $0$\\
$\alpha_1=L_1-\CO$&0&$\chi_1$&$0$&$0$\\
$\alpha_2=L_2-\CO$&0&$\chi_2$&$0$&$0$\\
$b=i_!\CO_E$&$0$&$0$  & $-\xi''_4$ & $0$\\
$c=i_{!}\CO_{p}$&$0$&$0$  & $0$ & $2\xi''_6$ \\
\hline
\end{tabular}
\end{center}
The line bundle $S$ on $W$ satisfies $\pi_{2}^{*}S=H^{5}$.
the ring structure is $a^{2}=25(b+c),\ ab=c$ and 
$a\alpha_{i}=b\alpha_{i}=0$.

The intersection form, in the above basis (omitting the 
$\alpha_{i}$), is
\begin{equation}
    \Omega=
    \begin{pmatrix}
    0&-5&0&-1\\
    5& 0& 1& 0\\
    0&-1& 0& 0\\
    1& 0& 0& 0
    \end{pmatrix}
\end{equation}
The monodromy about the large radius is generated by $M_{r=\infty}: v\mapsto 
v\otimes S$ or, in the above basis,
\begin{equation}\label{eq:MrW}
    M_{r=\infty}=
    \begin{pmatrix}
    1&0&0&0\\
    1&1&0&0\\
    0&25&1&0\\
    0&25&1&1
    \end{pmatrix}
\end{equation}

There are 5 conifold points, and the monodromies about them are 
generated by
\begin{subequations}\label{eq:McW}
    \begin{align}
         M_{c_{i}}&: v\mapsto v-(v,w_i)w_i,\qquad i=1,\dots4\\
         M_{c_{5}}&:v\mapsto v-25(v,\CO)\CO
    \end{align}
\end{subequations}
where
\begin{equation}
    \begin{split}
         w_1=&5\CO-4a+90b+6c\\
         w_2=&5\CO-3a+60b+8c\\
	 w_3=&5\CO-2a+35b+7c\\
	 w_4=&5\CO-a+15b+4c\\
    \end{split}
\end{equation}
These classes (including $k=5$, if one formally writes $w_5=5\CO$) can be 
written more succinctly as
\begin{equation}\label{eq:Wk}
    w_k=5\CO+(k-5)a+\tfrac{5}{2}(k-5)(k-10)b+\tfrac{1}{6}k(k-5)(k-10)c
\end{equation}

On the mirror of $W$, four of 
the five conifold points correspond to singularities at which
 an $S^3$ shrinks to zero size. At the mirror of the fifth,
there are five singularities, each of which locally looks like
$S^3/\BZ_5$. D-branes in such backgrounds were studied
in \cite{GopVaf}, with the result that five states should
become massless when the $S^3/\BZ_5$ shrinks to zero. This
result is consistent with the computation of the $F_1$-function
in \cite{AspMor}. The number of massless hypermultiplets
can be read off from the coefficient of the logarithm in an expansion 
of $F_1$ about the 
singular point. At $\psi=1$, there are 25 massless states. At the other four 
conifold points, there is only one massless state.

In our case, the multiplicity 25 in $M_{c_5}$ is easy to understand: as in 
\eqref{eq:sumL}, there are 25 6-branes (flat line bundles on $W$) which 
become massless at $\psi=1$. Near the other four points, the conformal 
field theory is the same as the theory near $\psi=1$, but with 
discrete torsion turned on. As we saw in \S\ref{sec:Wconf}, the 
remaining irreducible representations of the orbifold group on the Chan-Paton 
factors correspond to rank-5 projective bundles. But, since the 
discrete torsion is topologically trivial, we can choose an isomorphism 
between the twisted differential K-theory and the ordinary K-theory of 
$W$, and write these in terms of ordinary K-theory classes, $w_i$ on 
$W$. The rank of the $w_i$ are, of course, 5. The first Chern class (and
hence the coefficient of $a$) can be understood as follows.
The action
of a $2\pi$ rotation in the $\psi$-plane is to tensor with $S$, which 
shifts the first Chern class (and hence the coefficient of $a$) by 5.
A rotation
by $2\pi/5$ should therefore shift the first Chern class (and hence 
the coefficient of $a$) by 1.  Having 
understood these two coefficients, the rest are determined by 
requiring
\begin{equation*}
    w_{k+5}=w_k\otimes S
\end{equation*}
and requiring that the coefficients of $a,b,$ and $c$  vanish for $k=5$. The
unique solution to these constraints is \eqref{eq:Wk}.

In the above basis,
\begin{gather*}
 M_{c_{1}}= 
    \begin{pmatrix}
    -69 & -575 & -20 & -25\\
     56 &  461 &  16 &  20\\
  -1260 &-10350& -359&-450\\
    -84 & -690 &  -24& -29
    \end{pmatrix},\quad
 M_{c_{2}}=
    \begin{pmatrix}
	-34 & -425 & -15 & -25\\
	 21 &  256 &   9 &  15\\
       -420 &-5100 &-179 &-300\\
        -56 & -680 & -24 & -39
    \end{pmatrix},\\
 M_{c_{3}}=
    \begin{pmatrix}
	-14 & -300 & -10 & -25\\
	  6 &  121 &   4 &  10\\
       -105 &-2100 & -69 &-175\\
        -21 & -420 & -14 & -34
    \end{pmatrix},\quad
 M_{c_{4}}=
    \begin{pmatrix}
	-4 & -200 & -5 & -25\\
	 1 &   41 &  1 &   5\\
       -15 & -600 &-14 & -75\\
        -4 & -160 & -4 & -19
    \end{pmatrix},\\
 M_{c_{5}}=
    \begin{pmatrix}
         1 & -125 & 0 & -25\\
	 0 &    1 & 0 &   0\\
	 0 &    0 & 1 &   0\\
	 0 &    0 & 0 &1
     \end{pmatrix}
\end{gather*}

Multiplying these monodromies together, we find the monodromy at the 
Gepner point is 
\begin{equation}\label{eq:MidentityW}
    M_{g}= M_{c_{1}} M_{c_{2}} M_{c_{3}} M_{c_{4}} M_{c_{5}}M_{r=\infty}=
    1
\end{equation}
confirming that the Gepner point is a smooth point in the moduli 
space, as we argued in \S\ref{sec:Wconf}.

In contrast to the previous example, \eqref{eq:MrW} and \eqref{eq:McW} are not
the unique set of monodromies compatible with all of our criteria. For
instance, choose a particular torsion element, $\gamma\in K^0(W)_{tor}$, and
modify (\ref{eq:McW}a) to read
\begin{equation*}
   M_{c_{i}}: v\mapsto v-(v,w_i)(w_i+\gamma),\qquad
        i=1,\dots,4  \tag{\ref{eq:McW}a$'$}
\end{equation*}
This still acts trivially on the torsion subgroup, it induces the
\emph{same} action on $K^0(W)/K^0(W)_{tor}$ as before and, moreover, it still
satisfies
\eqref{eq:MidentityW} (because $\sum_{i=1}^4w_i=0$ mod 5).

Clearly more
physical input is required to completely pin down these monodromies in the
general case. That is an interesting subject for future work.

\section*{Acknowledgements}

We would like to thank
V.~Braun, E.~Diaconescu, M.~Douglas, B.~Fiol, G.~Moore,
T.~Pantev, C.~R\"omelsberger  and C.~Vafa
for discussions. J.D.~would also like to thank D.~Freed for invaluable
mathematical advice. Furthermore, we are grateful to the Erwin Schr\"odinger
Institute for hospitality during the workshop ``Duality, String Theory and
M-Theory,'' where some of this work was carried out.

\appendix

\bigskip
\section{Explicit Formulas for B-Type States on
Gepner Orbifolds.}
In this section, we provide explicit formulas for B-type
boundary states on Gepner orbifolds. For concreteness,
we include one orbifold generator $\gamma$, and a corresponding
lattice vector $\beta_1=(0; 2\gamma; 0,0,0,0,0)$. We assume
that $\gamma$ generates a free orbifold group of order $d= lcm\{k_i+2 \}$.
The closed string
partition function is \eqref{closedpf} 
where the lattice $\Lambda$ is generated by
$\beta_0, \beta^{(r)}, \beta_1$. 

B-type boundary states are written as a linear combination
of Ishibashi states  built on bulk fields, which fulfill
B-type boundary conditions. 
We impose B-type boundary conditions in each minimal model
separately, and therefore have to take bulk fields where
$\mu= -\overline{\mu}$. In addition, we have to require that
the bulk field is contained in the closed string partition function:
\begin{equation*}
\overline{\mu} = \mu + n\beta_0 + n_i\beta^{(i)} + m \beta_1 = -\mu.
\end{equation*}
In other words, all B-type Ishibashi states are built on winding modes
of the lattice $\Lambda$. 

Following Recknagel and Schomerus,
a consistent boundary state can then be obtained as
\begin{equation}
|\alpha\rangle= P \ \sum_{\lambda \mu} (-1)^{\frac{s_0^2}{2}} 
B_{\alpha}^{\lambda \mu} |\lambda \mu \rangle\rangle.
\end{equation}
Here, $\kket{\lambda \mu}$ is the Ishibashi state built on
the primary field of the representation $(\lambda, \mu)$,
$P$ is a projection operator which will be discussed in more
detail below and $B_{\alpha}^{\lambda\mu}$ is given by Cardy:
\begin{equation}
B^{\lambda,\mu}_{\alpha}\ =
        \prod_{j=1}^r{1\over{\sqrt{\sqrt{2}(k_j+2)}}}
                {{\sin(l_j,L_j)_{k_j}}\over
        {\sqrt{\sin(l_j,0)_{k_j}}}}\ e^{2\pi i \mu \sbullet \Xi}.
\end{equation}
The projector $P$ has to make sure that we sum over all winding
modes, and that the integer charge and spin alignment conditions
are fulfilled. 
\begin{multline}
P= \sum_{m\in\BZ_d} \ \sum_{n\in \BZ_{K/2}} \ 
\prod_{i=1}^{r} \  \frac{1}{k_i+2} \sum_{\rho_i\in \BZ_{k_i+2}}
e^{\frac{2\pi i}{k_i+2} \, (m_i + n + m \gamma_i) \, \rho_i} \\
\frac{1}{K} \sum_{\nu\in \BZ_K} e^{\pi i (2 \mu \sbullet \beta_0 -1)\nu} \ \ 
\prod_{i=1}^r \ \frac {1}{2} \ \sum_{\nu_i\in \BZ_2} 
e^{2\pi i \beta^{(i)} \sbullet \mu}
\end{multline}
The open string partition function computed from these boundary
states is given by
\begin{multline}
Z_{(\tilde \Lambda, \tilde M, \tilde M^1)(\Lambda, M, M^1)}
= const \ \sum_{\lambda' \mu'} 
\delta^{(d)}_{d \gamma\sbullet(\tilde \Xi-\Xi  - \mu')} \ \ 
\delta^{(K/2)}_{\sum \frac{K}{2k_i+4} (\tilde M_i -M_i - \nu -m_i')} \\
\prod_i N_{\tilde L_i L_i}^{l_i'} \ \delta^{(2)}_{\nu-M_i-\tilde M_i + m_i'}
 \ \ \delta^{(4)}_{\nu - 2\nu_i +S_i-\tilde S_i +s_i'}
\end{multline}
This is almost the same partition function as for B-type states on the
quintic, but there is an additional $\delta$-function insertion
in the open string sector (the first $\delta$-function in the above
equation). In words, this $\delta$-function enforces that all
momenta $\mu$ propagating in the open string sector are correctly
quantized along the direction $\beta_1$. 
(At this point it is easy to also describe boundary states on
orbifolds with more than one generator: The only
difference is that there are additional $\delta$-functions
imposing the correct momentum quantization 
in the open string partition function.)
The other $\delta$-functions 
make sure that the correct charge quantization is satisfied.
We read off from the partition function, that 
the boundary conditions are completely determined 
by $\Lambda$, $d\gamma\sbullet\Xi$
and $\sum M_i$, as noted in the text before. The quantities
$d\gamma\sbullet\Xi$ and  $\sum M_i$ appear as Wilson lines
in this formula: In the open string sector they determine a shift
in the momentum quantization along a particular direction. 
For the spectral flow operator $\beta_0$ this shift determines
if the two branes preserve the same supersymmetry. 
In \cite{Doug} this was related to a grading on the space of branes.
Here we see that the additional orbifold induces a similar kind
of grading. It would be interesting to interpret this further.

\bibliography{torsion}
\bibliographystyle{utphys}

\end{document}